\documentclass[12pt]{article}
\usepackage{amsfonts,amsthm,amsmath,amssymb,upgreek,bm}
\usepackage{hyperref}
\usepackage[paper=letterpaper,margin=.85in]{geometry}
\usepackage{graphicx}
\usepackage{units}
\usepackage{float}
\usepackage[export]{adjustbox}
\usepackage{xcolor}
\usepackage[shortlabels]{enumitem}
\usepackage[mathscr]{euscript}
\usepackage[normalem]{ulem}
\usepackage{bm}

\newcommand{\be}{\begin{equation}}
\newcommand{\ee}{\end{equation}}
\newcommand{\bea}{\begin{eqnarray}}
\newcommand{\eea}{\end{eqnarray}}

\renewcommand{\i}{\mathrm{i}}
\renewcommand{\d}{\mathrm{d}}

\renewcommand{\P}{\widehat{P}}
\newcommand{\p}{\widehat{p}}

\numberwithin{equation}{section}

\begin{document}
\thispagestyle{empty}

\vspace*{2.5cm}
\begin{center}

{\bf {\LARGE Firewalls from wormholes}}\\

\begin{center}

\vspace{1cm}

{\bf Douglas Stanford and Zhenbin Yang}\\
 \bigskip \rm

\bigskip 

Stanford Institute for Theoretical Physics,\\Stanford University, Stanford, CA 94305

\rm
  \end{center}

\vspace{2.5cm}
{\bf Abstract}
\end{center}
\begin{quotation}
\noindent

Spacetime wormholes can lead to surprises in black hole physics. 
We show that a very old black hole can tunnel to a white hole/firewall by  emitting a large baby universe. We study the process for a perturbed thermofield double black hole in Jackiw-Teitelboim (JT) gravity, using the lowest order (genus one) spacetime wormhole that corresponds to single baby-universe emission. The probability for tunneling to a white hole is proportional to $t^2 e^{-2S}$ where $t$ is the age of the black hole and $S$ is the entropy of one black hole.
\end{quotation}

\setcounter{page}{0}
\setcounter{tocdepth}{2}
\setcounter{footnote}{0}
\newpage

\parskip 0.1in
 
\setcounter{page}{2}
\tableofcontents

\newpage

\section{Introduction}
Stable black holes in AdS are expected to exhibit very small but erratic fluctuations in late time correlation functions $\langle \phi(t)\phi(0)\rangle$ \cite{Maldacena:2001kr,Dyson:2002nt,Barbon:2004ce}.
Polchinski suggested that these fluctuations might imply that old black holes have firewalls, because the erratic time dependence seems to conflict with fields being in the smooth infalling vacuum state near the horizon \cite{polchinskiPrivateCommunication}. Currently, a black hole explanation for the fluctuations in $\langle \phi(t)\phi(0)\rangle$ remains elusive, so it is hard to check Polchinski's suggestion directly. However, in some cases averages over these fluctuations have been explained using wormholes \cite{Saad:2018bqo,Blommaert:2019hjr,Saad:2019pqd,Yan:2022nod}, and one can ask: do such wormholes create firewalls? We will analyze this question in Jackiw-Teitelboim (JT) gravity and find that the answer is yes.

This relates to an open problem left in the wake of the Mathur/AMPS firewall paradox \cite{Mathur:2009hf,Almheiri:2012rt} and work addressing it \cite{Bousso:2012as,Nomura:2012sw,Verlinde:2012cy,Papadodimas:2012aq,Maldacena:2013xja,Penington:2019npb,Almheiri:2019psf,Almheiri:2019qdq,Penington:2019kki}: do firewalls exist for a black hole in a typical quantum state at some energy? Arguments that firewalls should exist were given in \cite{Almheiri:2013hfa,Marolf:2013dba}; other points of view include \cite{Susskind:2012rm,VanRaamsdonk:2013sza,Shenker:2013yza,Susskind:2015toa,deBoer:2018ibj,DeBoer:2019yoe,Susskind:2020wwe,Harlow:2021dfp,AhmedTalkACP}. One precise version of the question is as follows. Consider a state obtained by acting on the $L$ side of the thermofield double state with a simple operator $W$, and time evolving the other side by amount $t$:
\begin{align}\label{stateWt}
W_L\otimes e^{-\i H_R t} \ |\text{TFD}\rangle =\frac{1}{\sqrt{Z(\beta)}}\sum_{n} e^{-(\frac{\beta}{2}+\i t)E_n}W|n\rangle_L\otimes |\bar{n}\rangle_R.
\end{align}
By varying $t$ one can form a large set of states. If we choose $L,R$ to be black holes and pick a typical state from this set, is it safe to jump into $R$?

According to the classical black hole geometry, it is safe if $t > 0$, but it is dangerous if $t < 0$.\footnote{More precisely, the criterion should be $t \gtrless -t_*$ where $t_*$ is the scrambling time. In this paper we focus on timescales far longer than $t_*$ so we approximate $t_* = 0$.} The basic reason is illustrated below: the infaller from the black dot will not be affected by the $W$ perturbation if the time is positive, but it will suffer a collision if the time is negative.
\be
\includegraphics[valign = c, scale = 1.5]{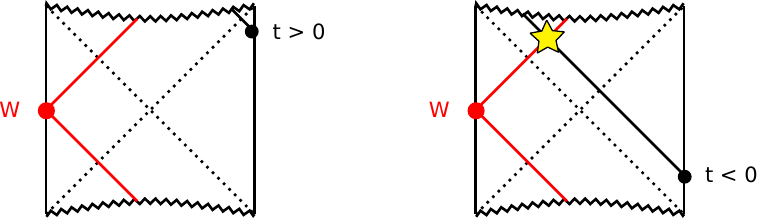}
\ee
This collision might look innocuous, but the relative boost between the $W$ perturbation and the infaller grows rapidly as $-t$ becomes large, and we will assume that this is effectively a firewall. (See appendix \ref{app:shock} for some discussion.) Another way of thinking about this \cite{Susskind:2015toa} is that for $t > 0$, we have a black hole and for $t < 0$, we have a white hole. It is safe to jump into black holes because their interiors are expanding and perturbations get diluted. But it is dangerous to jump into white holes, with contracting interiors and blueshifted perturbations. 

This classical answer -- ``safe for $t > 0$, dangerous for $t < 0$'' -- is not consistent with the finiteness of the black hole entropy. In finite entropy systems, the set of states one obtains by evolving far into the future is the same as the set one gets by evolving far into the past, and they should not have different properties for the infalling observer. Instead one might guess that if we evolve far into the future or the past, the quantum state will have a significant black hole component and also a significant white hole component, and that we end up with the same distribution for both signs of $t$. This scenario was referred to as a ``gray hole'' by Susskind \cite{Susskind:2015toa}.

In this paper, we explore a mechanism for the gray hole scenario, where black holes can tunnel into white holes by emitting a baby universe. This is a slight generalization of the ``wormhole shortening'' effect identified by Saad \cite{Saad:2019pqd} who showed in JT gravity  that the length of an Einstein-Rosen bridge can be reduced by a tunneling process that emits a baby universe.\footnote{JT gravity \cite{Teitelboim:1983ux,Jackiw:1984je} is a simple quantum gravity theory of two dimensional negatively curved spacetimes. Some references useful for this paper are \cite{Yang:2018gdb,Saad:2019lba,Saad:2019pqd}.} The tunneling process can be visualized as the following evolution of a spatial slice of the two-sided black hole:
\be\label{pinchingSketch}
\includegraphics[valign = c,scale = 1.5]{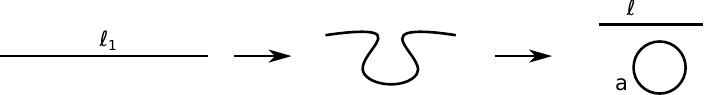}
\ee
On the left we start out with a long spatial slice of length $\ell_1$. Two points on the $\ell_1$ slice approach each other and pinch off, and we end up with a spatial slice of length $\ell$ and a closed universe of length $a$. There is an approximate conservation  in the total spatial length \cite{Saad:2019pqd}, so that $\ell \approx \ell_1- a$ -- emitting a baby universe makes the spatial wormhole shorter. 

Because the spatial length of a black hole interior is proportional to its age \cite{Hartman:2013qma,Susskind:2014rva,Yang:2018gdb}, one can also say that emitting the baby universe makes the black hole younger:
\be\label{effectiveAge}
(\text{effective age of BH after tunneling}) = (\text{age beforehand}) - (\text{size of baby universe}).
\ee
It turns out that this statement generalizes correctly to the case $a > \ell_1$, where the effective age after tunneling is negative. This case does not correspond to the cartoon (\ref{pinchingSketch}) directly, but it can be obtained by applying negative time evolution to both the $\ell_1$ and $\ell$ slices.

Eq.~(\ref{effectiveAge}) can be derived using the JT gravity trumpet path integral $Z_{\text{trumpet}}(\beta) = e^{-a^2/4\beta}/\sqrt{4\pi\beta}$. After replacing $\beta$ by $\beta + \i(t-t')$, the trumpet can be interpreted as a tunneling amplitude from the thermofield double state of age $t$ to the thermofield double state with age $t'$ plus a baby universe of size $a$:
\begin{align}
\langle \text{age $t'$};a|\text{age $t$}\rangle &= \includegraphics[valign = c, scale = 1]{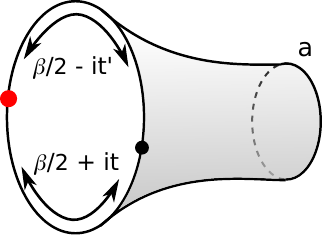}\\ &= \frac{\exp\left[-\frac{a^2}{4(\beta + \i (t-t'))}\right]}{\sqrt{4\pi(\beta + \i (t-t'))}}\\
&\approx \frac{\exp\left[\i\frac{a^2}{4 (t-t')} - \beta\frac{a^2}{4(t-t')^2}\right]}{\sqrt{4\pi \i (t-t')}}.
\end{align}
To study this amplitude at fixed energy, we multiply by $e^{\beta E}$ and integrate over $\beta$ along an inverse Laplace transform contour. This imposes a delta function that sets $2\sqrt{E}(t-t') = \pm a$. This reproduces (\ref{effectiveAge}) up to a sign that will be analyzed below.

For the states defined in (\ref{stateWt}), this means that even if $t$ is large and positive, baby universe emission allows the effective value of $t$ experienced by the infalling observer to be negative -- it might be a firewall.

The main task in this paper is to calculate the probability that the effective age of the black hole is negative, given that the original $t$ parameter is large and positive. We refer to this as $P_{\text{firewall}}(t)$. We work with JT gravity using \cite{Yang:2018gdb,Saad:2019pqd}. In the leading disk topology, there is no baby universe emission, and the probability of a firewall is zero for large $t$. At genus one, the tunneling process in (\ref{pinchingSketch}) can happen. As a function of the ``true'' age of the state, we find that the probability of tunneling to a negative ``apparent'' age is
\begin{align}\label{hdanswerintro}
P_{1,\text{firewall}}(t) &= \frac{1}{2}\left(\frac{t}{2\pi}\right)^2 e^{-2S(E)}.
\end{align}
Here the subscript $1$ refers to the genus, and $S(E)$ is the thermal entropy of a single black hole at energy $E$. This is our main result.

The important aspect of (\ref{hdanswerintro}) is the enhancement by $t^2$. This is related to the fact that the baby universe is characterized by a size and a twist $\{a,s\}$, and that these parameters need to be integrated over. We argue for a specific region over which they should be integrated, using the spatial slice that defines the interior to gauge-fix the mapping class group.

The growth with $t$ leads to two puzzles. First, how does the entire probability distribution remain normalized if the handle disk gives a contribution that grows with time? We argue that there are accompanying negative contributions that effectively borrow probability mass from the leading disk answer to pay for (\ref{hdanswerintro}). Second, what happens at $t \sim e^{S}$? We don't know, but an appealing possibility is that including the full genus expansion (and perhaps its nonperturbative completion) will lead to a symmetric late time value $P_{\text{firewall}} = P_{\text{smooth}} = \frac{1}{2}$.

In the rest of the paper we will compute (\ref{hdanswerintro}). Our strategy is to first compute the probability distribution for the length of the spatial slice across the interior of the black hole. Then we will go back one step in the computation and determine whether the spatial slice is expanding (black hole) or contracting (white hole). We illustrate this first with the disk topology.

\section{Disk}
In this section, we study the disk topology and compute $P_0(\ell)$, the probability distribution for the geodesic length $\ell$ across a two-sided black hole. The subscript zero refers to the genus. $P_0(\ell)$ depends on the age $t$ and energy $E$ of the black hole, and in the approximation of large energy $E$ and large $t$, the formula is very simple:
\be\label{quoted}
P_0(\ell) = \delta(\ell -\ell_t), \hspace{20pt} \ell_t = 2\sqrt{E}t -\log(4E).
\ee
The answer (\ref{quoted}) can be derived by computing the length of a geodesic in the classical geometry of the thermofield double black hole. However, we will practice for later sections by deriving it from the quantum formulation of JT gravity \cite{Yang:2018gdb,Kitaev:2018wpr,Saad:2019pqd}. 

\subsection{JT gravity wave function setup} The starting point is to decompose the path integral on the disk topology into an inner product of two states in the Hilbert space of JT gravity on an interval \cite{Yang:2018gdb,Kitaev:2018wpr,Harlow:2018tqv}, as in the following picture
\be
\includegraphics[valign =c, scale = 1.2]{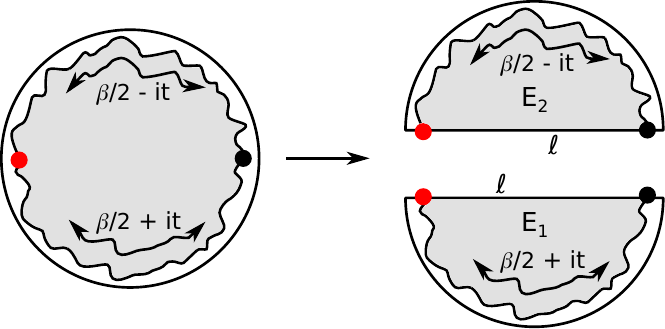}
\ee
The wiggly boundary represents the shape of the regularized boundary, with fluctuations described by the Schwarzian theory \cite{Jensen:2016pah,Maldacena:2016upp,engelsoy2016investigation}. In later drawings we will leave the regularized boundary and its wiggles implicit. The bottom half of this disk corresponds to the Hartle-Hawking state for boundary time $\frac{\beta}{2} + \i t$, expressed in the length basis:
\begin{align}
\langle \ell|\tfrac{\beta}{2}+ \i t\rangle &= \int_0^\infty \d E_1 \rho(E_1) \langle \ell|E_1\rangle \langle E_1|\tfrac{\beta}{2}+\i t\rangle\\
&=\int_0^\infty \d E_1 \rho(E_1) \langle \ell|E_1\rangle e^{-(\frac{\beta}{2}+\i t)E_1}.
\end{align}
Similarly, the top half represents $\langle \tfrac{\beta}{2}+\i t|\ell\rangle$. In JT gravity, the wave functions $\langle \ell|E\rangle$ \cite{Bagrets:2017pwq,Kitaev:2018wpr,Yang:2018gdb,Saad:2019pqd} and the density of states $\rho(E)$ \cite{Cotler:2016fpe,Stanford:2017thb} are.\footnote{Using the holographic regularization scheme: $\d s^2|_{\text{bdy}} = \frac{1}{\epsilon^2}\d u^2$, $\phi|_\text{bdy}={1\over 2\epsilon}$, and $\ell=\ell_{\text{actual}}+2\log \epsilon$.}
\be\label{JTgravAns}
\langle \ell|E\rangle = 2^{3/2} K_{2\i \sqrt{E}}(2e^{-\ell/2}), \hspace{20pt} \rho(E) = \frac{\sinh(2\pi \sqrt{E})}{(2\pi)^2}.
\ee
They satisfy the orthogonality and completeness relations
\begin{align}\label{orthogonality}
\int_{-\infty}^\infty \d \ell\, \langle E|\ell\rangle \langle \ell|E'\rangle &= \frac{\delta(E-E')}{\rho(E)},\\
\int_0^\infty \d E \rho(E) \,\langle \ell|E\rangle\langle E|\ell'\rangle &= \delta(\ell-\ell').
\end{align}
The disk partition function is obtained by gluing together two of these wave functions by integrating over $\ell$:
\begin{align}
Z_{\text{disk}} &=  e^{S_0}\int_{-\infty}^\infty \d \ell \langle \tfrac{\beta}{2}+\i t|\ell\rangle \langle \ell|\tfrac{\beta}{2}+\i t\rangle\\
&=  e^{S_0}\int_{-\infty}^\infty \d \ell  \int_0^\infty \d E_1 \d E_2 \rho(E_1) \rho(E_2) \langle E_2|\ell\rangle \langle \ell|E_1\rangle e^{-\frac{\beta}{2}(E_1 + E_2) - \i t (E_1 - E_2)}.
\end{align}
The result is exactly independent of $t$, since the integral over $\ell$ sets $E_1 = E_2$.

To get the probability distribution for the length $\P_0(\ell)$, one sets up the same computation but omits the integral over $\ell$. More precisely, we will study a fixed-energy version of the probability distribution, where we integrate over $\beta$ along an inverse Laplace transform contour:
\begin{align}
\P_0(\ell) \equiv e^{S_0}\int \frac{\d \beta}{2\pi \i}e^{\beta E}\langle \tfrac{\beta}{2}+\i t|\ell\rangle \langle \ell|\tfrac{\beta}{2}+\i t\rangle.
\end{align}
The integral over $\beta$ sets $E_1 + E_2 = 2E$, so we can parametrize the energies as
\be
E_1 = E + \sqrt{E}\omega, \hspace{20pt} E_2 = E - \sqrt{E}\omega,
\ee
and the probability distribution is
\begin{align}\label{prob}
\P_0(\ell) &= e^{S_0}\int (2\sqrt{E}\d\omega)\rho(E_1) \rho(E_2) \langle E_2|\ell \rangle\langle \ell|E_1\rangle e^{ - 2\sqrt{E}\i t \omega}.
\end{align}
We introduced the notation $\P$ instead of $P$ because of a technicality: the $\P$ is an un-normalized probability distribution, and to get the properly normalized $\P$ we will have to divide by the partition function.

In this paper we are interested in very long times $t$, and we can simplify the calculations a bit by implicitly ``smearing out'' $t$ and/or $\sqrt{E}$ by integrating these variables against smooth window functions. For $\sqrt{E}$, a small time-independent width is sufficient. For $t$, we should choose a width $1\ll \Delta t \ll t$. This will be useful in later computations, but one immediate consequence is that it will force $\omega$ to be small in the integral (\ref{prob}).

\subsection{Semiclassical approximation} We would like to evaluate (\ref{prob}) approximately, with the assumption that $E$ and $t$ are both large. For these purposes, we can use a semiclassical approximation to the wave functions $\langle \ell|E\rangle$. The semiclassical regime is large $E$, and a plot of the wave function looks like this:
\be\label{wfplot}
\includegraphics[valign = c, scale = .9]{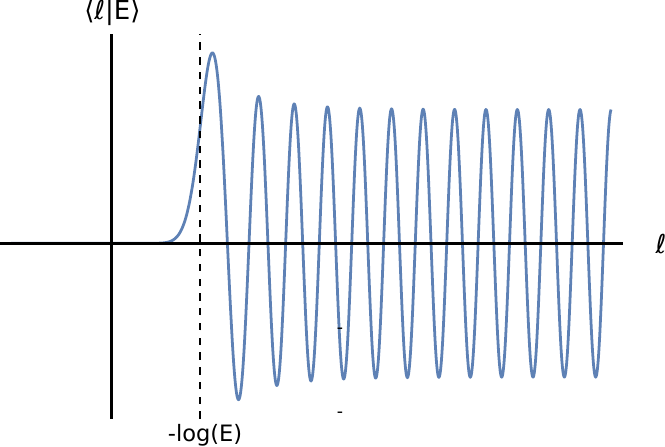} 
\ee
The oscillating region starts approximately at $\ell = -\log(E)$. This is the reglarized length of the shortest geodesic that connects the left and right boundaries of the thermofield double black hole. For smaller values of $\ell$, the wave function is exponentially suppressed, and for larger values, it oscillates.

In the oscillating region, the wave function (\ref{JTgravAns}) can be approximated as
\be\label{expanded}
\langle \ell|E\rangle \approx \frac{(8\pi)^{1/2}}{E^{1/4}} e^{-\pi \sqrt{E}}\cos\Big[\sqrt{E}(\ell +\log(4E)-2) - \frac{\pi}{4}\Big].
\ee
The cosine has two terms $e^{\pm\i \sqrt{E}\ell}$, corresponding to the expanding $(+)$ and contracting $(-)$ portions of the wave function. In the classical geometry of the thermofield double black hole, the expanding branch corresponds to $\ell$ being a spatial slice that passes through the future interior, and the contracting branch passes through the past interior (white hole). 

Because both the bra $\langle E_2|\ell\rangle$ and the ket $\langle \ell|E_1\rangle$ contain two branches, there are four total terms in integrand of (\ref{prob}). Expanding the phases to linear order in $\omega$ (which is appropriate if we smear out $t$ and/or $E$), these four terms have the following schematic form
\be
\left(\underbrace{e^{\i\omega \ell}}_{++} + \underbrace{e^{+2\i\sqrt{E}\ell}}_{-+} + \underbrace{e^{-\i\omega \ell}}_{--} + \underbrace{e^{-2\i\sqrt{E} \ell}}_{+-}\right) e^{-2\sqrt{E}\i t\omega}.
\ee
Here the $\pm\pm$ symbols refer to the branch choice for the bra and ket, respectively. Only the $++$ term will survive the integral over $\omega$, and it gives\footnote{The below is correct after integrating over $t$ or $\sqrt{E}$ with smooth window functions, as discussed above.}
\begin{align}
\P_0(\ell) &\approx e^{S_0}\frac{e^{2\pi\sqrt{E}}}{2(2\pi)^2}\int \frac{\d \omega}{2\pi}e^{\i \omega(\ell -\ell_t)}\\
&\approx e^{S(E)}\delta(\ell -\ell_t).\label{ledtodisk}
\end{align}
Here we defined
\be
\ell_t = 2\sqrt{E}t - \log(4E).
\ee
Recall that $\P$ is an un-normalized probability distribution. The properly normalized $P$ is obtained by dividing by the microcanonical partition function, $P = e^{-S(E)} \P$, giving (\ref{quoted}).

So far, we have computed the distribution for the geodesic length $\ell$ across a two-sided black hole, in the approximation of the disk topology. However, the computation also gives more than that: we found that only the expanding component of the wave function contributes. This means that we have a black hole and not a white hole. This could be formalized by working with the probability distribution of the relative boost angle of the two-sided black hole, but for simplicity we will just continue to work with $\ell$ and insert an extra step of keeping track of the expanding vs.~contracting branches.

\section{Handle disk}\label{sec:handle}
In this section, we keep the same boundary conditions as in the disk case, but we study a bulk topology with a handle. Our goal is to compute the probability distribution for the length of the spatial slice connecting two boundary points, keeping track of whether the spatial slice is expanding (black hole) or contracting (white hole). We are interested in the case where $e^S$ is large and $t$ is large, keeping terms in the probability distribution with mass of order $t^2 e^{-2S}$.

\subsection{First method}
To analyze the JT path integral, it is convenient to decompose the handle disk as follows:
\be\label{handlediskdecomp}
\includegraphics[valign = c, scale = 1.2]{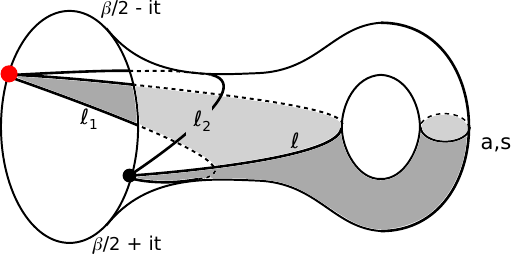}
\ee
Here $\ell$ is the spatial slice connecting the two boundary points, and $\ell_1,\ell_2$ are geodesics that separate off the asymptotic trumpet region. There is a closed universe of size $a$ and twist parameter $s$.

The shaded region in (\ref{handlediskdecomp}) corresponds to the tunneling process sketched in the introduction (\ref{pinchingSketch}). The JT gravity expression for this part of the geometry is \cite{Saad:2019pqd}
\begin{align}
\includegraphics[valign = c,scale = 1.1]{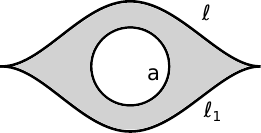} &\equiv \langle \ell,a|\ell_1\rangle\\[-10pt]
&=\int_0^\infty \d E \frac{\cos(a\sqrt{E})}{2\pi\sqrt{E}} \langle \ell|E\rangle \langle E|\ell_1\rangle\\
&= 2 K_0\left(2\sqrt{e^{-\ell} + e^{-\ell_1} + 2 e^{-(\ell + \ell_1)/2}\cosh\tfrac{a}{2}}\right).\label{eqn:BUemissionop}
\end{align}
Naively, we can form the handle disk geometry by gluing together two of these objects, and also gluing them to wave functions $\langle \tfrac{\beta}{2}+\i t|\ell_2\rangle$ and $\langle \ell_1|\tfrac{\beta}{2}+\i t\rangle$. However, this is subtle because for a given geometry, there are actually topologically distinct choices of $\ell$ (or $a$) geodesics, and a naive gluing sums over different geometries {\it and} different choices of geodesics. Below, we will see that this restricts the integration range of the $a$ and $s$ parameters. So for now we will leave the integration domain for $a,s$ unspecified.

The decomposition in (\ref{handlediskdecomp}) corresponds to
\begin{align}
\P_1(\ell) &\stackrel{?}{=} e^{-S_0}\int\frac{\d \beta}{2\pi\i}e^{\beta E}\int_{-\infty}^\infty \d \ell_1\d\ell_2\int \d a \d s \langle \tfrac{\beta}{2}+\i t|\ell_2\rangle\langle \ell_2|\ell,a\rangle\langle \ell,a|\ell_1\rangle\langle \ell_1|\tfrac{\beta}{2}+\i t\rangle\\
&=e^{-S_0}\int \d a \d s \int (2\sqrt{E}\d \omega) \langle E_2|\ell\rangle\langle \ell|E_1\rangle e^{-2\sqrt{E}\i t\omega}\frac{\cos(a\sqrt{E_1})}{2\pi\sqrt{E_1}}\frac{\cos(a\sqrt{E_2})}{2\pi\sqrt{E_2}}.
\end{align}
In going to the second line we inserted the energy representation to replace the integrals over $\ell_1$ and $\ell_2$ by integrals over $E_1,E_2$. The $\beta$ integral then fixed $E_1 + E_2 = 2E$, and we parametrized the difference of energies in terms of $\omega$ as in the disk computation:
\be
E_1 = E + \sqrt{E}\omega, \hspace{20pt} E_2 = E - \sqrt{E}\omega.
\ee
The product of cosines expands to four oscillating terms, and we keep the two terms where the phases approximately cancel if $E_1 \approx E_2$. Similarly, in the semiclassical approximation of the wave functions $\langle \ell|E_1\rangle$ and $\langle E_2|\ell\rangle$, one has four oscillating terms of which we keep the two where the phases approximately cancel (this corresponds to the expanding branches for the $\ell_1,\ell_2$ slices). One ends up with four terms total, corresponding to the independent choices of $\pm$ signs in the following expression (where we expanded the phase to linear order in $\omega$ using large $\Delta t$)
\begin{align}
\P_1(\ell)&\stackrel{?}{\approx} \frac{e^{-S(E)}}{2^5\pi^3E}\int \d a \d s\int\d\omega \exp\left\{\i\omega\left(\pm(\ell + \log(4E)) \pm a - 2\sqrt{E}t\right)\right\}\theta(\ell + \log E)\\
&\sim \frac{e^{-S(E)}}{2^4\pi^2E}\int\d a\d s \ \delta(\pm\ell \pm a - \ell_t)\,\theta(\ell).\label{deltafnin}
\end{align}
In the second line, we omitted some unimportant $\log(E)$ terms and did the integral over $\omega$.

The delta function and theta function in (\ref{deltafnin}) can be satisfied in three different ways $\ell = \ell_t\pm a$ and $\ell = a-\ell_t$. These three terms correspond to three different types of geometry that can contribute to $P(\ell)$. In the late-time limit that we are interested in, the lengths $\ell_1,\ell_2$ will be large. Also, for most of the support of the probability distribution, $\ell$ and $a$ will likewise be large. But the Gauss-Bonnet theorem implies that the area of the tunneling amplitude region remains small, so the geometry can be approximated by thin strips. In the three cases, the portion of the geometry between $\ell_1$ and $\ell_2$ can be approximated by the following thin strip geometries:
\be\label{Cases}
\includegraphics[valign = c, scale = 1.55]{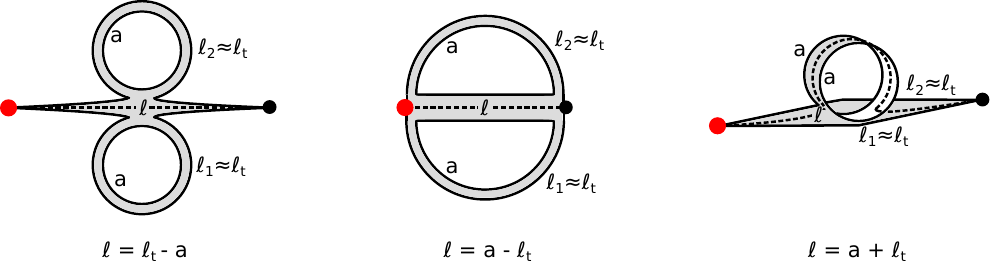}
\ee
In each case, the two boundaries labeled $a$ should be identified with each other, up to a twist by distance $s$. These geometries are a good approximation in the sense that distances between points are correct up to order one precision.

We can use these strip diagrams to confront the main subtlety, namely that different $\ell$ geodesics are possible on the same geometry. If our goal was to compute the partition function of the handle disk, we could pick any gauge-fixing condition to determine $\ell$, such as requiring it to be the shortest geodesic connecting the boundary points. But we are interested in the fate of the infalling observer, and we need to choose the $\ell$ slice that defines the correct interior. At least for genus one, it seems to be possible to select this using a very simple criterion: the spatial slice should not have shortcuts, meaning that distant points along the spatial $\ell$ slice should not be close together through the spacetime geometry. This criterion is designed to avoid overcounting geodesics that cover portions of the same fundamental spatial slice more than once. We will now discuss this criterion for the three cases (\ref{Cases}).

$\boldsymbol{ \ell = \ell_t - a}$: The delta function that is satisfied at this location comes from the integral over $\omega$ of $e^{\i\omega(\ell+a-\ell_t)}$, which arises from the term $e^{\i \sqrt{E_1}\ell}$ in the wave function $\langle \ell|E_1\rangle$. This corresponds to the expanding (firewall-free) branch of the wave functions. From the thin strip diagram, it is clear that for any value of $a$ and the twist $s$, distant points on the $\ell$ slice are not close to each other, so the ``no shortcut'' criterion does not impose any restriction on the moduli. The twist $s$ should therefore be integrated over the full fundamental region $0<s<a$, and $a$ should be integrated over $0<a<\ell_t$, where the upper limit comes from requiring that $\ell = \ell_t-a$ is positive. The total probability mass of this region is
\be
\P_{1,\text{smooth}} = \frac{e^{-S(E)}}{2^4\pi^2E}\int_0^{\ell_t} \d a\int_0^a\d s = \frac{e^{-S(E)}}{(2\pi)^2}{t^2\over 2}.
\ee

$\boldsymbol{\ell = a-\ell_t}$: This term arises from $e^{\i \omega(-\ell+a-\ell_t)}$, which corresponds to a contracting (firewall) branch of the wave functions. In this case, the ``no shortcut'' criterion is important. To see this, suppose first that the twist satisfies $0 < s < \ell$. Then for example the two points with $\mathsf{x}$ marks below will be identified with other:
\be
\includegraphics[valign = c, scale= 1.6]{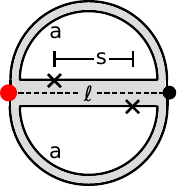} \label{HDfirewall}
\ee
Keeping in mind that the strip is very thin, this means that distant points along the $\ell$ geodesic are actually very close to each other in the full geometry. 

In this situation, there is a more fundamental geodesic $\ell'$ that is a better candidate for the correct spatial slice. Below we draw $\ell$ as the dashed geodesic, and $\ell'$ as the thick solid geodesic. The two drawings below are equivalent, related by an action of the mapping class group
\be\label{mcg1}
\includegraphics[valign = c,scale = 1.6]{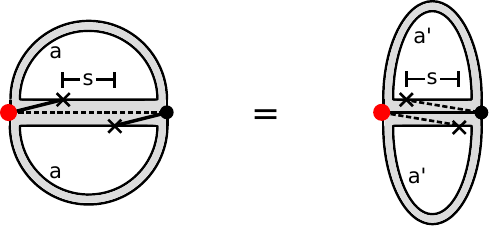}
\ee
The representation on the right makes it clear that the dashed $\ell$ geodesic is essentially a repetition of a portion of the solid $\ell'$. The mapping class group element that maps the left sketch into the right can be viewed as mapping $\ell$ to $\ell'$, and ``unwinding'' the $\ell$ geodesic once, removing a partial repetition of length $s$. In the thin strip approximation, it acts as (assuming $\ell > s$)
\be
(a,s,\ell) \to (a',s',\ell') = (a-s,s,\ell-s).
\ee
If we iterate this operation, we will eventually reach a situation with $\ell < s$ where there are no shortcuts. This is the geodesic we should use to define the interior without overcounting.

As an aside, we can ask what happens if we blindly act with the same element of the mapping class group once more, in the situation with $\ell < s$. The analog of (\ref{mcg1}) is now
\be
\includegraphics[valign = c,scale = 1.6]{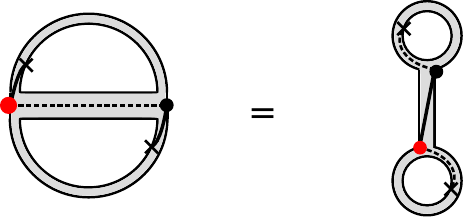}
\ee
and the action of the mapping class group element in the strip approximation is
\be
(a,s,\ell)\to (a',s',\ell') = (a-s,\ell,s-\ell).
\ee 
In this case, the $\ell$ and $\ell'$ geodesics are quite different from each other. The $\ell'$ geodesic might be shorter than the $\ell$ geodesic, or it might be longer, depending on the value of $s$. But because $\ell'$ does not correspond to any of the cases in (\ref{Cases}), it apparently should not contribute. The reason for this is that the $\ell'$ geodesic does not intersect the boundary orthogonally, and in the next section we will see that this means that the action is not stationary with respect to the positions of the endpoints. So the correct geodesic is sometimes but not always the shortest one.

The bottom line is that if $\ell > |s|$, the $\ell$ geodesic can be simplified to a more fundamental geodesic $\ell'$ of length $\ell - |s|$, but if $\ell < |s|$, it cannot. So in the moduli space integral, we should restrict the twist to satisfy $|s| > \ell$. Within the fundamental domain $0 < s < a$ for the twist, this condition becomes $s > \ell$ and $(a-s) > \ell$. Using the delta function that sets $\ell = a-\ell_t$, we can rewrite these constraints as
\be
a-\ell_t < s < \ell_t.
\ee
These inequalities imply $a < 2\ell_t$, and in order for $\ell$ to be positive, we need $a > \ell_t$. So the total  probability mass of this portion of the distribution is
\be\label{p1firewallmaintext}
\P_{1,\text{firewall}} = \frac{e^{-S(E)}}{2^4\pi^2E}\int_{\ell_t}^{2\ell_t}\d a \int_{a-\ell_t}^{\ell_t}\d s = \frac{e^{-S(E)}}{(2\pi)^2}{t^2\over 2}.
\ee
Multiplying by the normalization factor $e^{-S(E)}$ gives the answer quoted in the introduction (\ref{hdanswerintro}). 

Note that $P_{1,\text{firewall}} = P_{1,\text{smooth}}$ although this was not obvious at intermediate stages, because the computation of $P_{1,\text{smooth}}$ did not depend on any mapping class group issues. 

$\boldsymbol{\ell = \ell_t + a}$: This again corresponds to the expanding (firewall-free) branch of the wave functions. However, for this class of thin-strip geometry, because the $\ell$ geodesic follows the entire $a$ geodesic, any value of the twist $s$ will lead to a nonlocal identification along the $\ell$ geodesic. More precisely, the $\ell$ geodesic closely follows the $\ell_1$ or $\ell_2$ geodesics except for an excursion around the $a$ cycle. We interpret this to mean that the $\ell_1$ or $\ell_2$ geodesics define the correct interior. Computing the total probability mass in this region is subtle because the closed geodesic that is homologous to the asymptotic boundary (called $b$ below) is small, and the naive gluing procedure above does not include the mapping class group restriction on the corresponding twist parameter $\tau$. In the next section we will treat this more accurately and conclude that the small $b$ region gives a negative contribution that ``borrows'' probability from the disk answer.

\subsection{Second method}
The full handle disk path integral is an integral over the following variables:
\be
\text{(two points on asymptotic bdy)}, \text{(bdy wiggles connecting these pts)}, b,a,s
\ee
The integral over the boundary wiggles can be done by using the wave functions $\langle \frac{\beta}{2}+\i t|\ell_2\rangle$ and $\langle \ell_1|\frac{\beta}{2}+\i t\rangle$, and we are left with seven coordinates to integrate over: the locations of two points together with $b,a,s$. In the region between the $b$ geodesic and the asymptotic boundary, it is convenient to use the coordinates
\be
\d s^2 = \d \sigma^2 + \cosh^2\sigma \d\tau^2, \hspace{20pt} \tau \sim \tau + b
\ee
and to parametrize the boundary $(\sigma,\tau)$ points as $(\sigma_1,x_1)$ and $(\sigma_2,b/2 + x_2)$:
\be
\includegraphics[valign = c,scale = 1.3]{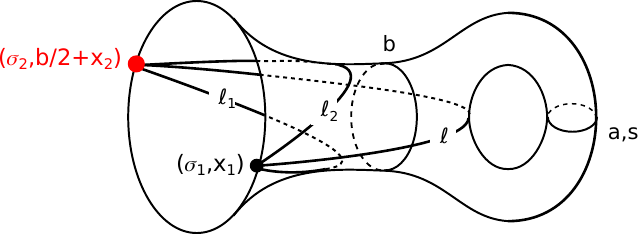}\label{fig:HD}
\ee
We would like to integrate over this seven dimensional space with the constraint that $\ell$ is held fixed, and subject to some inequalities that ensure $\ell$ defines the physical interior.

The variable $\sigma_1-\sigma_2$ does not affect the lengths $\ell,\ell_1,\ell_2$, so it can be integrated out, leaving a six-dimensional space that can be parametrized in terms of $\ell_1,\ell_2,b,\tau,a,s$ where $\tau$ is the twist on the $b$ cycle. In terms of these variables, the integrand consists simply of the wave functions $\langle \frac{\beta}{2}+\i t|\ell_2\rangle$ and $\langle \ell_1|\frac{\beta}{2}+\i t\rangle$ together with a factor of $\langle \ell_2,b|\ell_1\rangle$ that represents the remainder of the trumpet path integral. Note that this $\langle \ell_2,b|\ell_1\rangle$ is the same function that appeared in (\ref{eqn:BUemissionop}), as a baby universe creation operator, but here its interpretation is different. The only factor associated to the handle part of the spacetime is the flat Weil-Petersson measure $\d a \d s$:
\be
\P_1(\ell) = e^{-S_0}\int\frac{\d\beta}{2\pi\i}e^{\beta E}\,\d\ell_1\,\d\ell_2\, \d b\,\d\tau\, \d a\,\d s\, \langle \tfrac{\beta}{2}+\i t|\ell_2\rangle\langle\ell_2,b|\ell_1\rangle \langle \ell_1|\tfrac{\beta}{2}+\i t\rangle \delta(\ell - \ell(\text{moduli}))
\ee
Here $\ell$ is the argument of the distribution $P_1(\ell)$ that we trying to compute, and $\ell(\text{moduli})$ is the length of the geodesic labeled $\ell$ in the figure, see (\ref{Exactell}).

\subsubsection{Three-holed sphere with \texorpdfstring{$a > \ell_t$}{a>t}}
As a warmup for the handle disk computation, we will first consider the case where the handle is replaced by a three-holed sphere with two holes of fixed size $a$:
\be
\includegraphics[valign = c, scale = 1.2]{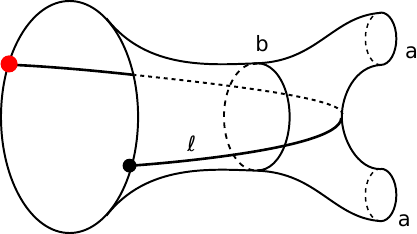}
\ee
We will focus on the case $a > \ell_t$ which contains the firewall part of the probability distribution, and we will refer to the length distribution as $\p(\ell)$:
\be
\p(\ell) = e^{-S_0}\int\frac{\d\beta}{2\pi\i}e^{\beta E}\,\d\ell_1\,\d\ell_2\, \d b\,\d\tau\, \langle \tfrac{\beta}{2}+\i t|\ell_2\rangle\langle\ell_2,b|\ell_1\rangle \langle \ell_1|\tfrac{\beta}{2}+\i t\rangle \delta(\ell - \ell(\text{moduli})).
\ee

Let's first discuss the integral $\int \d \ell \p(\ell)$. This integral removes the $\delta$ function, and one can then do the $\ell_1$ and $\ell_2$ integrals using orthogonality of the wave functions, getting\footnote{Smearing over $E$ a bit allows us to drop the contribution from $b = \infty$ below.}
\begin{align}
\int_{-\infty}^\infty \d \ell \p(\ell) &= e^{-S_0}\int\frac{\d\beta}{2\pi\i}e^{\beta E}\int_0^\infty\d b\int_0^{b/2}\d\tau\int \d E' \frac{\cos(b\sqrt{E'})}{2\pi\sqrt{E'}}e^{\beta E'}\\
&=e^{-S_0}\int_0^\infty b\d b \frac{\cos(b\sqrt{E})}{4\pi \sqrt{E}}\\
&= -e^{-S_0}\frac{1}{4\pi E^{3/2}}.\label{totalpl}
\end{align}
The twist $\tau$ was integrated from zero to $b/2$ instead of $b$ because of the $\pi$ rotation symmetry of the three-holed sphere. An important feature of this result is that the whole answer comes from the lower limit of the integral, near $b = 0$. This is true if we integrate over $\ell$, but for fixed $\ell$ there are important contributions from large $b$.

To calculate $\p(\ell)$ for fixed $\ell$, it is convenient to change variables
\be\label{cov}
\{\ell_1,\ell_2,b,\tau\} \to \{x_{12},\hat\sigma_+,b,\tfrac{x_1+x_2}{2}\},
\ee
where $x_{12} = x_1-x_2$, and $\hat \sigma_+ = \hat \sigma_1+\hat \sigma_2$ is the sum of the regularized $\sigma$ variables $e^{\hat \sigma}=\epsilon e^{\sigma}$. Changing variables from $\tau$ to $\frac{x_1+x_2}{2}$ is just a change of gauge for the overall $U(1)$ gauge symmetry. The change of variables from $\ell_1,\ell_2$ to $\hat\sigma_+,x_{12}$ follows from
\begin{align}
e^{\ell_1}=e^{\hat{\sigma}_+}\sinh^2({b\over 4}-{x_{12}\over 2}),\hspace{20pt} e^{\ell_2}=e^{\hat{\sigma}_+}\sinh^2({b\over 4}+{x_{12}\over 2}).\label{ell1ell2hd}
\end{align}
To simplify further, we will analyze the large $b$ region with $b_* < b < \infty$, where we have imposed a lower cutoff $b_*$ satisfying $1 \ll b_* \ll t$. After working out the answer from this region, we will use (\ref{totalpl}) to indirectly work out the contribution from small $b$. The simplification of the lower cutoff is that we can now take $b$ to be large. With this assumption, the measure in the new coordinates is simple $\d \ell_1\d\ell_2\d b = 2 \d \hat\sigma_+\d x_{12}\d b$, so 
\be\label{Pgeneral}
\p(\ell)=2e^{-S_0}\int \frac{\d\beta}{2\pi\i}e^{\beta E}\d x_1\d x_2\d\hat\sigma_+ \d b \langle \tfrac{\beta}{2}+\i t|\ell_2\rangle\langle \ell_1|\tfrac{\beta}{2}+\i t\rangle \langle \ell_2, b|\ell_1\rangle\delta(\ell-\ell(\text{moduli})).
\ee
Here and below, $\p(\ell)$ stands for the probability distribution for the length on the three-holed sphere geometry.

One can pull out the pieces that depend on $\beta$ and do the integral by imitating the steps that led to (\ref{ledtodisk}) in the disk computation
\begin{align}
\int \frac{\d\beta}{2\pi\i}e^{\beta E}\langle \tfrac{\beta}{2}+\i t|\ell_2\rangle\langle \ell_1|\tfrac{\beta}{2}+\i t\rangle &= \int (2\sqrt{E}\d \omega)\rho(E_1) \rho(E_2) \langle E_2|\ell_2 \rangle\langle \ell_1|E_1\rangle e^{ - 2\sqrt{E}\i t \omega}\\
&\approx \frac{e^{(2\pi+\i(\ell_1-\ell_2))\sqrt{E}}}{2(2\pi)^2}\int \frac{\d\omega}{2\pi} e^{\i \omega(\frac{\ell_1 + \ell_2}{2}-\ell_t)}\\
&= \frac{e^{(2\pi+\i(\ell_1-\ell_2))\sqrt{E}}}{2(2\pi)^2}\delta\left(\tfrac{\ell_1+\ell_2}{2}-\ell_t\right)\label{wle}\\
&\approx \frac{e^{2\pi\sqrt{E}}}{2(2\pi)^2} e^{-2\i \sqrt{E} x_{12}}\delta\left( \hat\sigma_++{b\over 2}-\ell_t\right).\label{delta111}
\end{align}
The integrand also contains the factor $\langle \ell_2, b|\ell_1\rangle$ which is given by (\ref{eqn:BUemissionop}):
\be
\langle \ell_2, b|\ell_1\rangle= 2 K_0\left(4 e^{-{\hat \sigma_+\over 2}}{\sinh{b\over 2}\over \cosh{b\over 2}-\cosh x_{12}}\right)\approx 2 K_0\left(4 e^{-{\hat \sigma_+\over 2}}\right)\approx \hat \sigma_+\theta(\hat \sigma_+).\label{werkj}
\ee
We used large $b$ in the first approximation. The second approximation is valid for large $|\hat\sigma_+|$, and although we will not use this below, it is helpful for intuition. In particular, the factor of $\hat \sigma_+$ can be understood as coming from the integral over $\hat\sigma_1-\hat \sigma_2$ which is implicit in $\langle \ell_2,b|\ell_1\rangle$ (see appendix \ref{app:buemission}).  Also, the theta function $\theta(\hat\sigma_+)$ forces $\hat\sigma_+$ to be positive, so $b$ is effectively confined to the range
\be
b<2\ell_t.\label{eqn:brange}
\ee
Inserting (\ref{delta111}) and (\ref{werkj}), (\ref{Pgeneral}) becomes
\be\label{pghj}
\p(\ell)\supset e^{-S_0}\frac{e^{2\pi\sqrt{E}}}{(2\pi)^2} \int_{b_*}^{\infty} \d b\, 2K_0(4e^{\frac{b-2\ell_t}{4}}) \int \d x_1\d x_2 e^{-2\i \sqrt{E} x_{12}}\delta\left(\ell-\ell(\text{moduli})\right)|_{\hat\sigma_+=\ell_t-{b\over 2}}.\ee

The general expression for $\ell(\text{moduli})$ is given in (\ref{Exactell}). For most of the support of $P_{\text{firewall}}$, we will have $a \gg \ell_t$, and together with (\ref{eqn:brange}) this implies $a \gg b/2$. This condition simplifies the general expression somewhat, to (\ref{approxell}):
\be\label{ellexpr}
 \ell(\text{moduli})= \hat \sigma_+ + a - \frac{b}{2} + 2\log 2 +  2\log\cosh\frac{x_1}{2} + 2\log\cosh\frac{x_2}{2} .
\ee
This expression for $\ell$ has an intuitive explanation using a strip diagram representation
\be\label{strip2}
\includegraphics[valign = c, scale= 1.3]{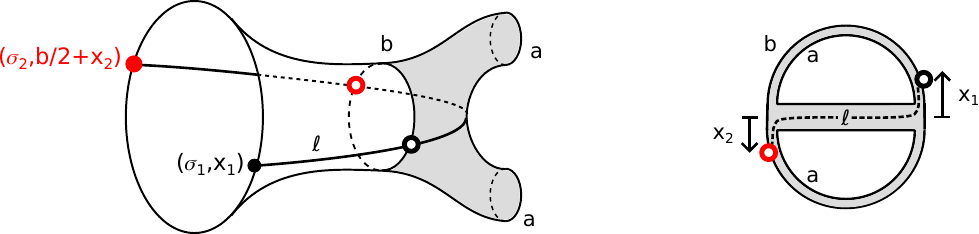}
 \ee
Here (in contrast to strip diagrams we drew earlier), the strip corresponds only to the shaded region of the geometry at left in (\ref{strip2}), with the outer circle representing the $b$ geodesic. As before, the inner loops represent two copies of the $a$ geodesic. The portion of the $\ell$ curve shown on the strip diagram has length $a-\frac{b}{2}+|x_1|+|x_2|$. In the full formula (\ref{ellexpr}) the $|x|$ functions are smoothed out into $2\log\cosh\frac{x_1}{2}$, and we also have a $\hat\sigma_+$ term, which represents the distance from the regularized asymptotic boundary to the $b$ circle.

The integral (\ref{pghj}) is now completely explicit
\be
\begin{split}
\p(\ell)& \supset  e^{-S_0}{e^{2\pi \sqrt{E}}\over (2\pi)^2} \int_{b_*}^{\infty} \d b \, 2K_0(4e^{\frac{b-2\ell_t}{4}}) 
\int_{-\infty}^\infty \d x_1 \d x_2e^{-2\i \sqrt{E}x_{12}} \\
& \hspace{30pt}\times\delta\left(\ell-(\ell_t + a - b + 2\log 2 +  2\log\cosh\frac{x_1}{2} + 2\log\cosh\frac{x_2}{2})\right)
\end{split}
\ee
In principle, the $x$ variables should be integrated over a finite range proportional to $b$, but the large $b$ assumption allows us to replace this with the whole real line.\footnote{This is a bit subtle because endpoint contributions in oscillating integrals are numerically large. But they will be oscillating as a function of $b$, and after using the $\delta$ function constraint, they will be oscillating as a function of $\ell$ and $\ell_t$, and they will be suppressed by smearing out $t$ and/or $\sqrt{E}$.} Because of the oscillating phase factor $e^{-2\i \sqrt{E} x_{12}}$, the dominant contribution comes from the small $x$ region where the $\log\cosh (x/ 2)$ functions are close to a nonanalyticity. In other words, the integral over the boundary wiggles that led to this phase factor effectively imposes an on-shell condition that $\ell$ should be extremal with respect to the angular variables. In this region, we have $\ell \approx \hat \sigma_++a-{b\over 2}=\ell_t+a-b$.

To do the $x_1,x_2$ integrals, we can expand the $\delta$ function in powers of the $\log \cosh (x_i/ 2)$ quantities. The important term is the lowest-order term where both $\log\cosh (x_i/2)$ variables appear:
\be\label{goodterm}
\delta(\ell-\ell(\text{moduli}))=\dots+4\log\cosh\frac{x_1}{2} \log\cosh\frac{x_2}{2} \delta''(\ell+b-\ell_t-a)+ \dots
\ee
This is the important term because lower order terms will give zero after integrating over $x_i$, and higher-order terms will be proportional to sufficiently high derivatives of $\delta$ functions that the resulting contribution to the probability distribution will locally integrate to zero. The $x_i$ integral of the term involving (\ref{goodterm}) can be analyzed using
\begin{align}
\int_{-\infty}^\infty\d x \, e^{-2\i\sqrt{E}x}\log(\cosh\frac{x}{2}) &= -\frac{\pi}{2\sqrt{E}\sinh(2\pi\sqrt{E})} \approx -\frac{\pi}{\sqrt{E}}e^{-2\pi\sqrt{E}}.\label{xintegral}
\end{align}
Keeping only the term (\ref{goodterm}) and doing the $x$ integrals with (\ref{xintegral}), we get
\be
\int \d x_1 \d x_2e^{-2\i \sqrt{E}x_{12}} \delta(\ell-\ell(\text{moduli}))\approx {4\pi^2\over E} e^{-4\pi\sqrt{E}}\delta''(\ell+b-\ell_t-a)
\ee
and therefore
\begin{align}
\p(\ell) &\supset \frac{e^{-S(E)}}{8\pi^2 E}\int_{b_*}^{\infty} \d b \, 2K_0(4e^{\frac{b-2\ell_t}{4}}) \delta''(\ell+b-\ell_t-a)\\
&=\frac{e^{-S(E)}}{8\pi^2 E}\left[\partial_\ell^2 2K_0(4e^{-\frac{\ell+\ell_t-a}{4}})+(\text{terms from $b_*$ limit})\right].\label{eqn:3hs}
\end{align}
The second derivative of the Bessel function is a positive function that is localized near the region $\ell+\ell_t-a = 0$. Since we are implicitly smearing out $t$ and/or $E$ with window functions, we can approximate this Bessel expression as a delta function with the same total weight
\be
 \partial_\ell^2 2K_0(4e^{-\frac{\ell+\ell_t-a}{4}}) \to \frac{1}{2}\delta(\ell+\ell_t-a).
\ee

Comparing to our earlier computation, this term reproduces the ``firewall'' delta function in (\ref{deltafnin}). However, what the current method makes clear is that there are additional terms coming from the small $b$ region, and that they must have negative total integral so as to reproduce (\ref{totalpl}). We interpret these extra terms as interference terms for which the correct spatial slice is $\ell_1$ or $\ell_2$, and not the slice $\ell$. We will try to justify this interpretation in the next paragraphs, but first let's work out its implication. The $\ell_1$ and $\ell_2$ slices have length close to $\ell_t$, so if we let $L$ be the length of the physical spatial slice, the total distribution will be
\be\label{physicalL}
\p(L) = \frac{e^{-S(E)}}{2^4\pi^2 E}\left[\delta(L+\ell_t-a) - \delta(L-\ell_t)\right] -\frac{e^{-S_0}}{4\pi E^{3/2}}\delta(L-\ell_t).
\ee
Here we have simply used that the small $b$ terms must contribute something proportional to $\delta(L-\ell_t)$, with a coefficient determined by (\ref{totalpl}).

We have three indirect arguments for the above interpretation. The first is based on consistency of the probability interpretation. Negative terms are not allowed in a probability distribution, but because the negative terms in (\ref{physicalL}) are proportional to $\delta(L-\ell_t)$, they can combine with the disk answer (\ref{ledtodisk}) to give a positive distribution. From this perspective it is also important to note that the $\ell_1$ and $\ell_2$ slices are expanding, so the negative terms in (\ref{physicalL}) represent expanding slices as in the disk geometry. This interpretation would not work if we had continued to view $\ell$ as the physical slice, becuase in the small $b$ region, we have $\ell \sim \ell_t + a$ which is different from the disk.

A second argument is the analogy to a model discussed in appendix \ref{app:defects} where the holes are replaced with non-backreacting defects. There, the positive term corresponds to an inner product of bulk states with one defect in the ket and one in the bra, meaning that the spatial slice goes between them and therefore resembles $\ell$. The negative terms correspond to bulk configurations with both defects in the ket or both in the bra, so that the spatial slice is more like $\ell_1$ or $\ell_2$.

A third, somewhat vaguer argument is based on the geometry with large $b$ vs.~small $b$. In the large $b$ region, we have $\hat\sigma_+ \approx 0$, so that the asymptotic boundary can be thought of as being close to the $b$ geodesic. A spatial geodesic fired orthogonally from the boundary will prefer to go through the holes. In the small $b$ region, we have $\hat\sigma_+ \approx \ell_t$, and the (regularized) angle (\ref{eqn:angle}) between the $\ell_1$ and $\ell_2$ slices becomes very small. The $\ell,\ell_1,\ell_2$ geodesics are very similar to each other, but the $\ell$ geodesic adds an extra loop around the $a$ cycle, creating a shortcut.

\subsubsection{Back to the handle disk}
Now we return to the case of the handle disk, where it is also necessary to integrate over $a$ and $s$. Compared to the three-holed sphere, the new subtlety is that we need to work out a restricted integration region that takes into account the ``no shortcut'' rule for the $\ell$ geodesic.

After identifying the two $a$ loops together in (\ref{strip2}), this strip geometry can be represented as a trivalent band geometry, made from three strips of lengths $y_1,y_2,y_3$, glued with a twist so that the geometry has a single boundary.
 \be\label{trivalentbandfig}
\includegraphics[scale = 1.5,valign = c]{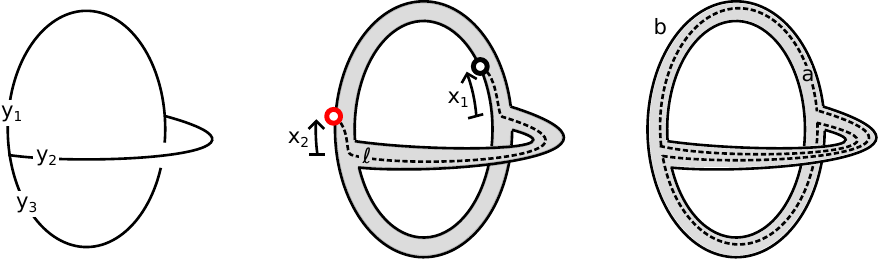}
 \ee
 At left, we labeled the lengths of three strips. In the center, we thickened the strips and indicated the $x$ locations of the boundary points, projected to the $b$ geodesic which forms the boundary of the strip. The length of this $b$ geodesic is
 \be
b = 2(y_1+y_2+y_3).
 \ee
At right, we showed the $a$ geodesic, which has length
 \be
a = y_1 + 2y_2+y_3.
 \ee
 This geodesic looks a bit odd, but it is determined by the condition that it should be a simple closed geodesic that does not intersect $\ell$. Note that points along the $a$ geodesic with separation $y_1+y_2$ are actually close together in the strip geometry. This determines
 \be
s = y_1 + y_2.
 \ee
These three relations can be inverted to get
 \begin{align}
y_1 = \frac{b}{2}-(a-s),\hspace{20pt}  y_2 = a-\frac{b}{2}, \hspace{20pt}
y_3 = \frac{b}{2}-s.
 \end{align}

Now, the condition we would like to impose is that the $\ell$ geodesic should not have shortcuts. This means that on the trivalent band geometry, portions of the $\ell$ geodesic should not overlap. So, for the case $x_1>0,x_2>0$ shown in (\ref{trivalentbandfig}), we require $y_1 > x_1+x_2$. Together with $y_3 > 0$ this leads to the bounds $a + x_1+x_2-b/2 < s < b/2$, so the total volume of the $s$ integral is $b - a - x_1-x_2$. By similarly analyzing cases with one or both of $x_1,x_2$ negative, one finds the formula
\be
\int_{\text{no shortcuts}}\d s = b - a - |x_1|-|x_2| = (\ell_t - \ell)\theta(\ell_t-\ell).
\ee
In the last step we assumed that the $|x|$ functions from the strip approximation should be replaced by their $2\log\cosh(x/2)$ counterparts, and we then used (\ref{ellexpr}) and the delta function in (\ref{delta111}).

We can use this to integrate the positive term from the three-holed sphere answer over the restricted moduli space of $a,s$
\begin{align}
\P_{1,\text{firewall}}(\ell) &= \int_{\ell_t}^\infty \d a (\ell_t-\ell)\theta(\ell_t-\ell)\frac{e^{-S(E)}}{8\pi^2E}\frac{1}{2}\delta(\ell+\ell_t-a)\\
&={e^{-S(E)}\over 16 \pi^2E}(\ell_t-\ell)\theta(\ell)\theta(\ell_t-\ell).
\end{align}
After integrating over $\ell$ we recover (\ref{p1firewallmaintext}). For the negative terms, the correct spatial slice is either the $\ell_1$ or $\ell_2$ slice, not $\ell$. So we should not impose the no shortcut condition for $\ell$. We will not try to analyze the negative terms in detail, because we believe they must effectively subtract from the disk answer a probability mass equal to that of the positive terms. To give a concrete formula, what we expect is that after combining the disk answer with the firewall and smooth genus one contributions, the probability distribution for the physical length should be
\begin{align}
P(L)&=\left(1 - \frac{t^2e^{-2S(E)}}{(2\pi)^2}\right)\delta(L- \ell_t) + \frac{e^{-2S(E)}}{8\pi^2 E}(\ell_t-L)\theta(\ell_t-L)\theta(L) + O(e^{-4S}).\label{Pdist}
\end{align}
The first term represents the contribution of the disk, together with the putative negative terms from the handle disk. The second term represents $P_{1,\text{firewall}}$ plus an identical term for $P_{1,\text{smooth}}$.

Finally, we would like to compare our computation here to the volume of the moduli space $V_{1,1}(b)$ \cite{naatanen1998weil,mirzakhani2007simple}. First, let's review the computation of that volume in the thin-strip limit. The trivalent band picture makes this computation simple, since in the $y_1,y_2,y_3$ coordinates, the mapping class group reduces to cyclic permutations which can be accounted for by ignoring the MCG and then dividing by three \cite{naatanen1998weil}. The measure can be obtained by computing
\be
\d (\frac{b}{2}) \d a \d s = \d y_1\d y_2\d y_3,
\ee
so we have
\be
V_{1,1}(b) = \frac{1}{3}\int_0^\infty \delta(y_1+y_2+y_3 - \frac{b}{2})\d y_1 \d y_2\d y_3 = \frac{1}{3}\cdot\frac{b^2}{8} = \frac{b^2}{24}.
\ee
This is the right answer for the volume in the large $b$ limit.\footnote{The volume $V_{1,1}(b)$ is often given as one half this value, because the genus one surface has a rotation symmetry that means that the twist on the $b$ cycle should only be integrated from zero to $b/2$, rather than $b$. By replacing $V_{1,1}(b)$ by half of its value here, one can then integrate the twist from zero to $b$.}

In our computation, we did not divide by the factor of three for the cyclic permutations of the $y_1,y_2,y_3$ lengths, and correspondingly our $a$ and $s$ integral was (with $x_1 = x_2 = 0$ and $b = 2\ell_t$)
\be
\int_{b/2}^b\d a \int^{b/2}_{a-b/2} \d s  = \frac{b^2}{8},
\ee
which is three times larger than the volume of the moduli space. This can be traced to the fact that there are three ``on shell'' choices of the $\ell$ geodesic -- corresponding to $\ell$ threading any one of the $y_1,y_2,y_3$ strips. These three choices intersect the $b$ geodesic at different points, so they correspond to placing the operators on the asymptotic boundary at different angles relative to the handle. Since we integrate over the locations of the operators on the asymptotic boundary, we should sum over these three. By choosing the $\ell$ geodesic to pass through the $y_2$ strip, we have used this sum to gauge-fix the cyclic $\mathbb{Z}_3$ mapping class group in the trivalent band picture.

\section{Discussion}

So is the horizon of an old black hole safe?

We found that it can be dangerous. The old black hole might have tunneled to a white hole, with a firewall on the horizon. The contribution to this probability from a genus one wormhole, $P_{1,\text{firewall}}(t)$, grows with time and becomes of order one when the age of the black hole is order $e^{S(E)}$. The wormhole that describes this tunneling also contributes to late-time two point functions \cite{Saad:2019pqd}, but in the present case it is enhanced by an additional zero mode integral that corresponds to the age of the final black hole in the tunneling process.\footnote{The idea that $e^{S_{BH}}$ final states can enhance tunneling was discussed in \cite{Mathur:2008kg} in the context of fuzzballs \cite{Mathur:2008nj}.} In the two point function, this final age is restricted by the operator insertions to be near zero, see appendix \ref{app:correlator}.

It is interesting to compare to the normalized spectral form factor (SFF)
\be
|\langle \text{TFD}|e^{-\i H_R t}|\text{TFD}\rangle|^2 = \frac{Z(\beta+\i t)Z(\beta-\i t)}{Z(\beta)^2}.
\ee 
After replacing $t \to t-t'$, this is the overlap of a  TFD state at time $t$ with one at time $t'$. In this paper, we computed the projection of the late-time TFD onto states with {\it bulk} age $t'$, defined with respect to a special spatial slice. In the calculation of $P_{1,\text{smooth}}$ this distinction isn't important, and our quadratic answer is precisely the integral of the linear ``ramp'' in the microcanonical spectral form factor between $0 < t' < t$. In systems without time-reversal, this linear ramp continues all the way to the plateau time $t=2\pi e^{S(E)}$, and it is interesting that extrapolation of $P_1$ to this time would lead to $P_{\text{firewall}} = P_{\text{smooth}}={1\over 2}$. So, perhaps the physics of the plateau could be relevant at late times.

What does the ``bulk age'' mean in the boundary theory? A related problem is to interpret the length of the Einstein-Rosen bridge. This was conjectured to be dual to the computational complexity of the boundary state \cite{Susskind:2014rva}. It was pointed out in \cite{Iliesiu:2021ari} that higher genus corrections will be important in computing the length at late time. In \cite{Iliesiu:2021ari}, a definition of the length was was used that involves a regularized sum over all possible geodesics. Our work suggests that just one geodesic could be used. From (\ref{Pdist}), we get
\be
\int\d \ell P(\ell)\ell = 2\sqrt{E}\left(t -\frac{t^3}{6\pi^2}e^{-2S(E)}\right) + O(e^{-4S(E)}).
\ee
As in \cite{Iliesiu:2021ari}, there are sizable corrections when $t$ is only a fraction of $e^{S(E)}$, which may not be consistent with expectations for computational complexity.\footnote{This was pointed out to us by L.~Susskind, but see \cite{Lin:2018cbk}.}  Another perspective is Krylov complexity \cite{Parker:2018yvk,Rabinovici:2020ryf,Balasubramanian:2022tpr} which would define bulk age by orthogonalizing time-evolved TFD states.

{\bf A concrete open problem} is to determine the value that $P_{\text{firewall}}(t)$ saturates to, by analyzing higher order effects. At very late times, a time-evolved state should become typical within the set of all possible time-evolved states, and among this class of states one would expect equal probability for a black hole or a white hole, suggesting $P_{\text{firewall}}(\infty) = 1/2$. It is not clear what notion of typicality is relevant, or how long it takes to set in, but the growth of $P_{1}(t)$ suggests that {\it something} has to happen around $t\sim e^{S(E)}$, and saturation to $1/2$ is a natural guess. The only hint of this within our calculations was that the genus one contribution (by itself) had equal probability of being a firewall or a smooth state.

{\bf A second open problem} is how our conclusions would change in a bulk theory dual to a specific quantum system, rather than an ensemble average. The half-wormhole model \cite{Saad:2021rcu,Saad:2021uzi,Blommaert:2021fob} suggests that we replace the handle disk by the drawing at left below, where the jagged boundaries represent pseudorandom boundary conditions associated to the specific theory:
\be
\includegraphics[valign = c, scale = 1.2]{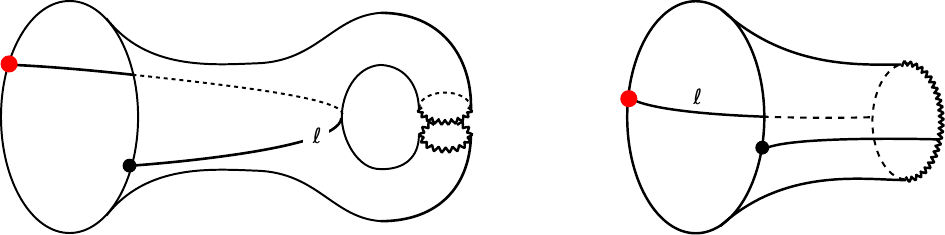}
\ee
Presumably the handle disk computation we did can be viewed as an average over the half-wormhole contributions at left. But we can also imagine a half-wormhole geometry like one on the right. Can this represent a firewall? This figure does not seem to contribute to the black hole $\to $ white hole tunneling, which is the firewall mechanism we examined in this paper. However, there might be more than one way to create a firewall: tunneling to a white hole is one way, but perhaps tunneling to a dangerous half-wormhole state is another.

\section*{Acknowledgements} 

We thank Ahmed Almheiri, Luca Iliesiu, Daniel Jafferis, Jorrit Kruthoff, Phil Saad, Arvin Shahbazi-Moghaddam, Stephen Shenker, Leonard Susskind and Shunyu Yao for discussions. DS is supported in part by DOE grant DE-SC0021085 and by the Sloan Foundation. ZY is supported in part by the Simons Foundation.

\appendix

\section{Shocks make the singularity close to the horizon}\label{app:shock}
For a white hole of age $t'<0$, classical geometry predicts that the infaller and the $W$ perturbation will collide with a center of mass energy of order
\be
E_{\text{c.o.m}}^2 \sim s \propto e^{\frac{2\pi}{\beta}(-t')}.
\ee
For the white holes that are discussed in this paper, with ages $t'$ of order $-e^{S(E)}$, this energy would be $\exp(\# e^{S(E)})$, a ridiculously large value. What should the infaller expect?

First, let's discuss scattering at more modest energies $s\,G_N \sim 1$, so that the scattering is approximately elastic. Then the infaller's experience can be described as propagation on a black hole with a shock wave near the horizon. This shock wave geometry is obtained by gluing together two halves of the unperturbed thermofield double geometry by a null translation by amount
\be
\Delta v \sim e^{\frac{2\pi}{\beta}(t - t_*)}
\ee
along the $u = 0$ horizon \cite{Dray:1984ha,Shenker:2013pqa}. Here $t_*$ is the scrambling time. The geometry with the shock is pictured at right in the Kruskal diagrams below:
\be
\includegraphics[valign = c, scale = .8]{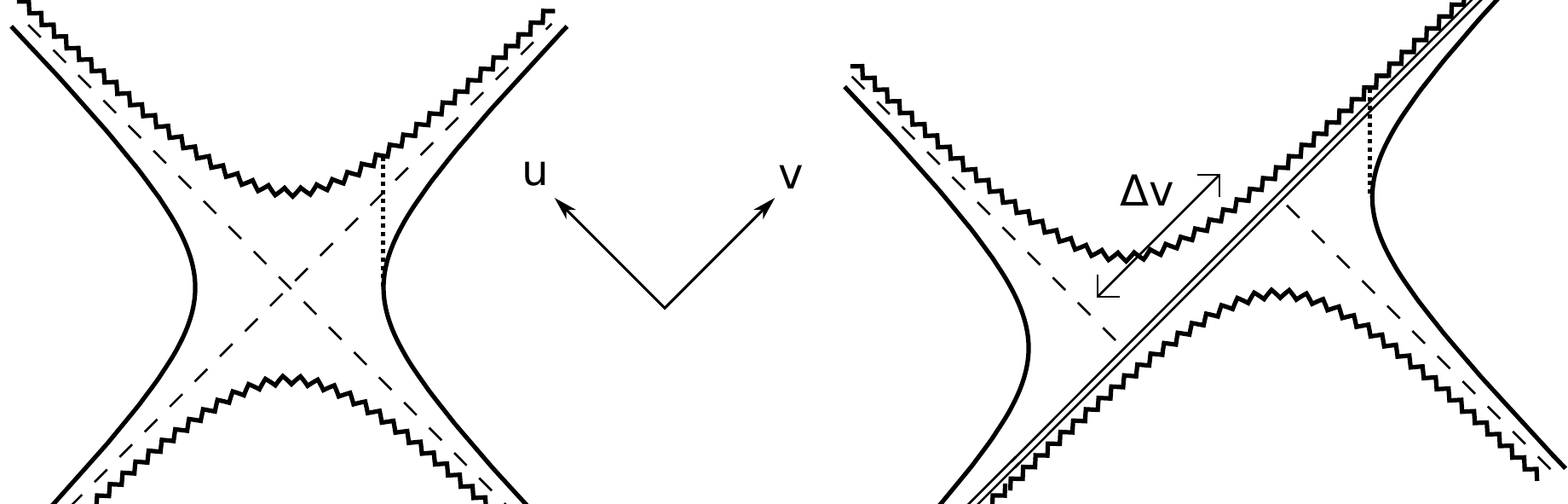}
\ee
We have also drawn an infaller's trajectory with dotted lines. It is apparent that the proper time the infaller experiences between the horizon and the singularity is reduced by the shock. For large $\Delta v$, one finds
\be
t_{\text{proper}} \propto \frac{1}{\Delta v} \sim e^{-\frac{2\pi}{\beta}(-t'-t_*)}\hspace{20pt} (-t' > t_*).
\ee
This means that for strong shocks (but still within the validity of this approximation) the singularity approximately coincides with the horizon. A plausible scenario is that the correct extrapolation of this experience to arbitarily high energies is simply that the infaller's experience effectively terminates at the horizon, and we do not need to fully answer the question of what happens in ultra-high energy scattering. Of course, it is also possible that there are surprises in store at very high energies and that this simple extrapolation is incorrect.\footnote{The tunneling from a black hole to a white hole or vice versa discussed in this paper can be considered such a surprise, but it is already taken into account for the purposes of this appendix.}

In this paper we have worked with JT gravity. In this case gravitational scattering is elastic at all energies, and there is no black hole singularity. The analog of the statement that the proper time until the singularity becomes small is that the infaller will experience a very rapidly changing dilaton after passing through the shock. In near extremal black holes to which JT is an approximation, there are singularities that the infaller will rapidly hit. We are not focusing too much on the aspects that are special to JT, because we hope that the overall picture in this paper, that the late-time black hole can tunnel into a white hole, will also be true in more interesting theories of quantum gravity.

\section{The length of \texorpdfstring{$\ell$}{L} on the handle disk}\label{app:detailsHandleDisk}
The purpose of this appendix is to compute the length $\ell$ of the geodesic connecting the red and black boundary points in the geometry below:
\be\label{trumpet3hole}
\includegraphics[valign = c, scale = 1.2]{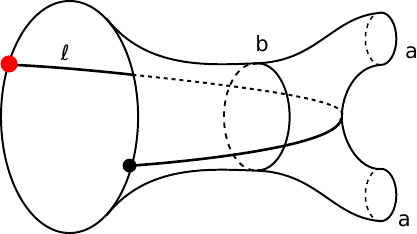}
\ee
This geometry consists of a trumpet with closed geodesic of length $b$ glued to a three-holed sphere with geodesic holes of lengths $a,a,b$. The endpoints of the geodesic $\ell$ are at the boundary of the trumpet region. In the trumpet region, we will use the coordinates
\be
\d s^2 = \d \sigma^2 + \cosh^2(\sigma)\d \tau^2, \hspace{20pt} \tau \sim \tau + b.
\ee
The three holed sphere with a trumpet attached has a $\mathbb{Z}_2$ reflection symmetry, and the fixed locus of the reflection is a special geodesic labeled $\gamma$ in the figure below. It is convenient to choose the origin of the $\tau$ coordinate so that the endpoints of $\gamma$ are at $\tau = 0$ and $\tau = b/2$.
\be
\includegraphics[valign = c, scale = 1.2]{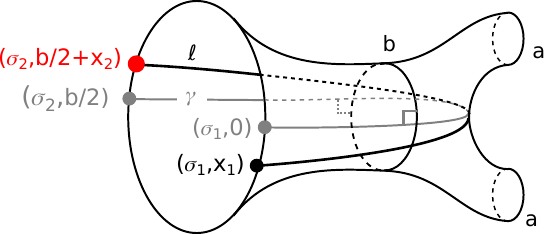}
\ee
We have used $x_1$ and $x_2$ to denote the separation in the $\tau$ coordinate between the $\ell$ geodesic (which we are trying to the compute the length of) and the special $\gamma$ geodesic. 

The exact formula for $\ell$ is
\begin{align}\label{Exactell}
e^{\ell} 
=\left(\cosh{x_1-x_2\over 2}+\cosh{x_1+x_2\over 2}\sqrt{1+{\sinh^2{b\over 4}\over \cosh^2{a\over 2}}}\right)^2 {\cosh^2{a\over 2}\over \sinh^2{b\over 4}}e^{\hat\sigma_1+\hat\sigma_2}.
\end{align}
It can be derived by representing the geometry (\ref{trumpet3hole}) as a quotient of the full hyperbolic space.

Intermediate steps in the derivation of (\ref{Exactell}) are quite complicated, so we will just check it in the limit $a \gg b/2 \gg 1$ that was used in the main text. In this limit, the length of the portion of $\gamma$ that lies within the three-holed sphere becomes long and (thanks to Shunyu Yao for the following formula, which was worked out using chapter 2 of \cite{thurston1979geometry})
\begin{align}
\text{length of $\gamma$ within three-holed sphere} &= 2\text{arccosh}\sqrt{1 + \frac{\cosh^2\frac{a}{2}}{\sinh^2\frac{b}{4}}}\\
&\approx a - \frac{b}{2} + 2\log 2.
\end{align}
This immediately allows us to compute the length of $\gamma$ itself, as 
\begin{align}
\text{length}(\gamma) &= \sigma_1 +\sigma_2 + \text{length within three-holed sphere}\\
&\approx \sigma_1 + \sigma_2 + a - \frac{b}{2} + 2\log 2.
\end{align}
Because the length of $\gamma$ within the three-holed sphere is long, the $\gamma$ and $\ell$ geodesics will closely approximate each other for most of this length. So we can compute the $x_1$ dependence by evaluating the $x_1$ dependence of a simpler problem: a geodesic that connects $(\sigma_1,x_1)$ to a distant point with $\tau$ coordinate equal to zero:
\be
\includegraphics[valign = c, scale = 1]{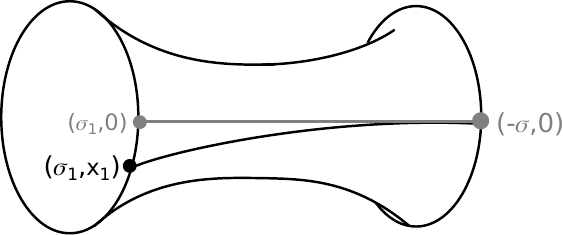}
\ee
For large $\sigma_1+\sigma$, this geodesic has length 
\be
\sigma_1 + \sigma + 2\log \cosh\frac{x_1}{2}
\ee
so we conclude that the $x_1$ dependent part of $\ell$ is $2\log\cosh\frac{x_1}{2}$. The logic is similar for the $x_2$-dependent part, and we conclude that (after holographic regularization)
\be\label{approxell}
\ell \approx \hat\sigma_1 +\hat \sigma_2 + a - \frac{b}{2} + 2\log 2 +  2\log\cosh\frac{x_1}{2} + 2\log\cosh\frac{x_2}{2} .
\ee
This agrees with (\ref{Exactell}) if $a \gg b/2 \gg 1$.

\section{Geometry of the baby universe emission amplitude}\label{app:buemission}
In this appendix, we will derive a simple expression of the baby universe emission amplitude $\langle \ell_2,b|\ell_1\rangle$ (\ref{werkj}) in terms of the trumpet coordinates of the two boundary points $(\sigma_{1,2},\tau_{1,2})$.
Since JT gravity only has boundary degrees of freedom, the action of this amplitude is purely geometrical, given by its renormalized hyperbolic area. 
Using the Gauss-Bonnet theorem, the hyperbolic area is given by the jump angles $\alpha_{1,2}$ at the two boundary points:
\be
A=-{1\over 2}\int R=-2\pi+\sum_{i}\alpha_i.
\ee
So we only need to work out the jump angles $\alpha_{1,2}$. 
It is most straightforward to work in the embedding coordinates $\vec Y=(Y^{-1},Y^{0},Y^{1})$:
\bea
&Y^{-1}=\cosh\sigma \cosh\tau,~Y^0=\cosh\sigma\sinh\tau,~Y^1=\sinh\sigma,&\\
&Y\cdot Y=\eta_{ab} Y^a Y^b=-(Y^{-1})^2+(Y^{1})^2+(Y^{1})^2=-1.&
\eea
In this coordinate, any geodesic is specified by a normal vector $\vec Z$ such that:
\be
\vec Z\cdot \vec Y=0.
\ee
In particular, given two points $A,B$ on the geodesic $\vec Y_A,\vec Y_B$, we have:
\be
\vec Z_{AB}={\vec Y_A \times \vec Y_B\over |\vec Y_A \times \vec Y_B|}.
\ee
Now the jump angle between two geodesics $AB$ and $CA$ intersecting at point $A$ is equal to the angle between their normal directions:
\be
\alpha_A=\arccos (\vec Z_{AB}\cdot\vec Z_{CA}).
\ee 
To get the jump angle $\alpha_2$ at the location $(\sigma_2,\tau_2) $ in (\ref{fig:HD}), we can use:
\be
\vec{Y}_2=\vec{Y}(\sigma_2,\tau_2), \vec{Y}_1=\vec{Y}(\sigma_1,\tau_1),\vec{Y}_{1'}=\vec{Y}(\sigma_1,b+\tau_1).
\ee
This gives:
\be
\alpha_2=4\epsilon {(e^b-1)e^{\tau_1+\tau_2-\hat\sigma_2}\over (e^{b+\tau_1}-e^{\tau_2})(e^{\tau_2}-e^{\tau_1})}=4\epsilon{\sinh{b\over 2}\over \cosh{b\over 2}-\cosh x_{12}} e^{-\hat\sigma_2},\label{eqn:angle}
\ee
where we have used $\tau_1 =x_1, \tau_2= {b\over 2}+x_2$. In the limit of $b\gg 1$, this simplifies:
\be
\alpha_2\sim 4 \epsilon e^{-\hat \sigma_2},
\ee
A similar analysis gives:
\be
\alpha_1\sim 4\epsilon e^{-\hat \sigma_1}.
\ee
Together we have area:
\be
A\sim -2\pi +4\epsilon (e^{-\hat \sigma_1}+e^{-\hat \sigma_2}).
\ee
The JT action is given by the renormalized area\cite{Yang:2018gdb}:
\be
I=-\phi_b (A+2\pi)=-4\phi_r(e^{-\hat \sigma_1}+e^{-\hat \sigma_2})=2e^{-\hat \sigma_1}+2e^{-\hat \sigma_2}.
\ee
where in the last step we used convention of this paper $\phi_r={1\over 2}$. The baby universe emission amplitude (\ref{werkj}) is
\be
\frac{1}{2}\int_{-\infty}^\infty \d \hat\sigma_- e^{-I}.
\ee

\section{Model with light defects}\label{app:defects}
In this appendix we will discuss a toy model that was helpful for us in interpreting calculations in this paper. Consider JT gravity with a gas of conical defects of angle $2\pi - \alpha$, as considered in \cite{Witten:2020ert,Maxfield:2020ale}. In the limit of small $\alpha$, the defects do not backreact, and they become marked points that should be integrated with measure 
\be\label{measure}
\lambda\alpha e^{-\alpha \phi(x)}\sqrt{g(x)}\d^2 x,
\ee
where $\lambda \ll 1$ is a fugacity that controls the expansion.

In the limit of small $\alpha$, path integrals with defect insertions are very easy to compute, because the factor of $\alpha$ ensures that the only contribution can come from the region where the insertion points explore the large volume near the boundary, and the result is therefore proportional to the regularized length of the boundary. Explicitly, near the boundary
\be
\phi = \frac{\phi_r}{z}, \hspace{20pt} \d s^2 = \frac{\d z^2 + \d u^2}{z^2}
\ee
and the integral over the small $z$ region gives
\begin{align}
\lambda\int \d u \int_{0}^z\frac{\d z'}{z'^2} \alpha e^{-\frac{\phi_r}{z'}\alpha} &= \frac{\lambda}{\phi_r}\int \d u\, e^{-\frac{\phi_r}{z}\alpha}\approx \frac{\lambda}{\phi_r}\int \d u.
\end{align}
where in the last line we used small $\alpha$. In the convention $\phi_r = 1/2$, this leads to the simple general formula
\be
[\text{path integral with bdy length $u$ and $k$ defects}] = \frac{(2\lambda u)^k}{k!}[\text{same thing w/o defects}].
\ee

Let's now get to the point of this model. We consider the time-evolved thermofield double state, $|\frac{\beta}{2}+\i t\rangle$ and let $|k\rangle$ be the $k$-defect contribution to it:
\be
|k\rangle = |\tfrac{\beta}{2}+\i t \text{ with $k$ defect insertions}\rangle.
\ee
Suppose that we want to compute the norm of the portion of the state that contains one defect in its preparation, namely
\begin{align}
\langle 1|1\rangle &= Z_{\text{disk}}\times (2\lambda)^2(\tfrac{\beta}{2}+\i t)(\tfrac{\beta}{2}-\i t)\\
&\propto t^2.
\end{align}
To compute this term, it is not sufficient to simply compute the two-defect (order $\lambda^2$) correction to the inner product $\langle \tfrac{\beta}{2}+\i t|\tfrac{\beta}{2}+\i t\rangle$. The reason is that the $\lambda^2$ term in this inner product contains two types of terms,
\begin{align}
\langle \tfrac{\beta}{2}+\i t|\tfrac{\beta}{2}+\i t\rangle|_{\lambda^2} &=\langle 1|1\rangle + \Big[\langle 0|2\rangle  + \langle 2|0\rangle\Big] \\ &= (2\lambda)^2 Z_{\text{disk}}\times\left\{(\tfrac{\beta}{2}+\i t)(\tfrac{\beta}{2}-\i t) + \Big[\frac{1}{2}(\tfrac{\beta}{2}+\i t)^2 + \frac{1}{2}(\tfrac{\beta}{2}-\i t)^2\Big]\right\}\\
&=(2\lambda)^2 Z_{\text{disk}}\times \left\{\frac{\beta^2}{4} + t^2 + \Big[\frac{\beta^2}{4}-t^2\Big]\right\}\\
&= (2\lambda)^2 Z_{\text{disk}}\times\frac{\beta^2}{2}.
\end{align}
The term $\langle 1|1\rangle$ is present, and it makes a positive contribution that grows with time. But there is also an interference-type term that comes with a negative sign and cancels the time-dependent part of $\langle 1|1\rangle$. This cancellation is necessary for unitarity, which implies that $\langle \tfrac{\beta}{2}+\i t|\tfrac{\beta}{2}+\i t\rangle$ should be independent of time.

\section{Connection to the correlator}\label{app:correlator}
In this appendix, we explain a connection to the correlation function computed by Saad in \cite{Saad:2019pqd}. The two-sided correlation function in the TFD at inverse temperature $\beta$ or energy $\sqrt{E} = \pi/\beta$ is proportional to 
\be
\frac{1}{\cosh^{2\Delta}(\frac{\pi t}{\beta})} = \frac{1}{\cosh^{2\Delta}(\sqrt{E}t)}.
\ee
As shown in the introduction, emitting a baby universe of size $a$ can change the age to
\be
\sqrt{E}t' = \sqrt{E}t - \frac{a}{2},
\ee
so one finds an answer proportional to (after normalizing by the disk)
\be
e^{-2S}\int \d a \d s \frac{1}{\cosh^{2\Delta}(\sqrt{E}t - a/2)}.
\ee
For large $\Delta$, the $a$ parameter will be localized near $2\sqrt{E} t$, and the only zero mode is the twist $s$, which gives a factor of $\int \d s = a \sim t$. However, if we take $\Delta$ to be small, then fluctuations in $a$ become large and the answer becomes of order
\be
\frac{t}{\Delta}e^{-2S}.
\ee
The factor of $t$ represents the $s$ zero mode and the factor of $1/\Delta$ represents fluctuations in $a$ around the value $2\sqrt{E}t$. These fluctuations have both positive and negative $t'$ and therefore include both black hole and white hole states.

For most of the integration region leading to this factor $t/\Delta$, the geodesic that is relevant for the correlator also contributes to the probability distribution we computed in this paper. In principle, the correlator includes a sum over geodesics, but for large $a$, this sum is not significant.

\bibliography{references}

\providecommand{\href}[2]{#2}\begingroup\raggedright\begin{thebibliography}{10}

\bibitem{Maldacena:2001kr}
J.~M. Maldacena, ``{Eternal black holes in anti-de Sitter},''
  \href{http://dx.doi.org/10.1088/1126-6708/2003/04/021}{{\em JHEP} {\bfseries
  04} (2003) 021},
\href{http://arxiv.org/abs/hep-th/0106112}{{\ttfamily arXiv:hep-th/0106112
  [hep-th]}}.

\bibitem{Dyson:2002nt}
L.~Dyson, J.~Lindesay, and L.~Susskind, ``{Is there really a de Sitter/CFT
  duality?},'' \href{http://dx.doi.org/10.1088/1126-6708/2002/08/045}{{\em
  JHEP} {\bfseries 08} (2002) 045},
  \href{http://arxiv.org/abs/hep-th/0202163}{{\ttfamily arXiv:hep-th/0202163}}.

\bibitem{Barbon:2004ce}
J.~Barbon and E.~Rabinovici, ``{Long time scales and eternal black holes},''
  \href{http://dx.doi.org/10.1002/prop.200410157}{{\em NATO Sci. Ser. II}
  {\bfseries 208} (2006) 255--263},
  \href{http://arxiv.org/abs/hep-th/0403268}{{\ttfamily arXiv:hep-th/0403268}}.

\bibitem{polchinskiPrivateCommunication}
J.~Polchinksi.
\newblock Private communication, Sept.~26 2017.

\bibitem{Saad:2018bqo}
P.~Saad, S.~H. Shenker, and D.~Stanford, ``{A semiclassical ramp in SYK and in
  gravity},''
\href{http://arxiv.org/abs/1806.06840}{{\ttfamily arXiv:1806.06840 [hep-th]}}.

\bibitem{Blommaert:2019hjr}
A.~Blommaert, T.~G. Mertens, and H.~Verschelde, ``{Clocks and Rods in
  Jackiw-Teitelboim Quantum Gravity},''
  \href{http://dx.doi.org/10.1007/JHEP09(2019)060}{{\em JHEP} {\bfseries 09}
  (2019) 060}, \href{http://arxiv.org/abs/1902.11194}{{\ttfamily
  arXiv:1902.11194 [hep-th]}}.

\bibitem{Saad:2019pqd}
P.~Saad, ``{Late Time Correlation Functions, Baby Universes, and ETH in JT
  Gravity},''
\href{http://arxiv.org/abs/1910.10311}{{\ttfamily arXiv:1910.10311 [hep-th]}}.

\bibitem{Yan:2022nod}
C.~Yan, ``{Crosscap Contribution to Late-Time Two-Point Correlators},''
  \href{http://arxiv.org/abs/2203.14436}{{\ttfamily arXiv:2203.14436
  [hep-th]}}.

\bibitem{Mathur:2009hf}
S.~D. Mathur, ``{The Information paradox: A Pedagogical introduction},''
  \href{http://dx.doi.org/10.1088/0264-9381/26/22/224001}{{\em Class. Quant.
  Grav.} {\bfseries 26} (2009) 224001},
  \href{http://arxiv.org/abs/0909.1038}{{\ttfamily arXiv:0909.1038 [hep-th]}}.

\bibitem{Almheiri:2012rt}
A.~Almheiri, D.~Marolf, J.~Polchinski, and J.~Sully, ``{Black Holes:
  Complementarity or Firewalls?},''
  \href{http://dx.doi.org/10.1007/JHEP02(2013)062}{{\em JHEP} {\bfseries 02}
  (2013) 062},
\href{http://arxiv.org/abs/1207.3123}{{\ttfamily arXiv:1207.3123 [hep-th]}}.

\bibitem{Bousso:2012as}
R.~Bousso, ``{Complementarity Is Not Enough},''
  \href{http://dx.doi.org/10.1103/PhysRevD.87.124023}{{\em Phys. Rev.}
  {\bfseries D87} no.~12, (2013) 124023},
\href{http://arxiv.org/abs/1207.5192}{{\ttfamily arXiv:1207.5192 [hep-th]}}.

\bibitem{Nomura:2012sw}
Y.~Nomura, J.~Varela, and S.~J. Weinberg, ``{Complementarity Endures: No
  Firewall for an Infalling Observer},''
  \href{http://dx.doi.org/10.1007/JHEP03(2013)059}{{\em JHEP} {\bfseries 03}
  (2013) 059},
\href{http://arxiv.org/abs/1207.6626}{{\ttfamily arXiv:1207.6626 [hep-th]}}.

\bibitem{Verlinde:2012cy}
E.~Verlinde and H.~Verlinde, ``{Black Hole Entanglement and Quantum Error
  Correction},'' \href{http://dx.doi.org/10.1007/JHEP10(2013)107}{{\em JHEP}
  {\bfseries 10} (2013) 107},
\href{http://arxiv.org/abs/1211.6913}{{\ttfamily arXiv:1211.6913 [hep-th]}}.

\bibitem{Papadodimas:2012aq}
K.~Papadodimas and S.~Raju, ``{An Infalling Observer in AdS/CFT},''
  \href{http://dx.doi.org/10.1007/JHEP10(2013)212}{{\em JHEP} {\bfseries 10}
  (2013) 212},
\href{http://arxiv.org/abs/1211.6767}{{\ttfamily arXiv:1211.6767 [hep-th]}}.

\bibitem{Maldacena:2013xja}
J.~Maldacena and L.~Susskind, ``{Cool horizons for entangled black holes},''
  \href{http://dx.doi.org/10.1002/prop.201300020}{{\em Fortsch. Phys.}
  {\bfseries 61} (2013) 781--811},
\href{http://arxiv.org/abs/1306.0533}{{\ttfamily arXiv:1306.0533 [hep-th]}}.

\bibitem{Penington:2019npb}
G.~Penington, ``{Entanglement Wedge Reconstruction and the Information
  Paradox},''
\href{http://arxiv.org/abs/1905.08255}{{\ttfamily arXiv:1905.08255 [hep-th]}}.

\bibitem{Almheiri:2019psf}
A.~Almheiri, N.~Engelhardt, D.~Marolf, and H.~Maxfield, ``{The entropy of bulk
  quantum fields and the entanglement wedge of an evaporating black hole},''
\href{http://arxiv.org/abs/1905.08762}{{\ttfamily arXiv:1905.08762 [hep-th]}}.

\bibitem{Almheiri:2019qdq}
A.~Almheiri, T.~Hartman, J.~Maldacena, E.~Shaghoulian, and A.~Tajdini,
  ``{Replica Wormholes and the Entropy of Hawking Radiation},''
  \href{http://dx.doi.org/10.1007/JHEP05(2020)013}{{\em JHEP} {\bfseries 05}
  (2020) 013}, \href{http://arxiv.org/abs/1911.12333}{{\ttfamily
  arXiv:1911.12333 [hep-th]}}.

\bibitem{Penington:2019kki}
G.~Penington, S.~H. Shenker, D.~Stanford, and Z.~Yang, ``{Replica wormholes and
  the black hole interior},'' \href{http://arxiv.org/abs/1911.11977}{{\ttfamily
  arXiv:1911.11977 [hep-th]}}.

\bibitem{Almheiri:2013hfa}
A.~Almheiri, D.~Marolf, J.~Polchinski, D.~Stanford, and J.~Sully, ``{An
  Apologia for Firewalls},''
  \href{http://dx.doi.org/10.1007/JHEP09(2013)018}{{\em JHEP} {\bfseries 09}
  (2013) 018}, \href{http://arxiv.org/abs/1304.6483}{{\ttfamily arXiv:1304.6483
  [hep-th]}}.

\bibitem{Marolf:2013dba}
D.~Marolf and J.~Polchinski, ``{Gauge/Gravity Duality and the Black Hole
  Interior},'' \href{http://dx.doi.org/10.1103/PhysRevLett.111.171301}{{\em
  Phys. Rev. Lett.} {\bfseries 111} (2013) 171301},
  \href{http://arxiv.org/abs/1307.4706}{{\ttfamily arXiv:1307.4706 [hep-th]}}.

\bibitem{Susskind:2012rm}
L.~Susskind, ``{Singularities, Firewalls, and Complementarity},''
  \href{http://arxiv.org/abs/1208.3445}{{\ttfamily arXiv:1208.3445 [hep-th]}}.

\bibitem{VanRaamsdonk:2013sza}
M.~Van~Raamsdonk, ``{Evaporating Firewalls},''
  \href{http://dx.doi.org/10.1007/JHEP11(2014)038}{{\em JHEP} {\bfseries 11}
  (2014) 038}, \href{http://arxiv.org/abs/1307.1796}{{\ttfamily arXiv:1307.1796
  [hep-th]}}.

\bibitem{Shenker:2013yza}
S.~H. Shenker and D.~Stanford, ``{Multiple Shocks},''
  \href{http://dx.doi.org/10.1007/JHEP12(2014)046}{{\em JHEP} {\bfseries 12}
  (2014) 046}, \href{http://arxiv.org/abs/1312.3296}{{\ttfamily arXiv:1312.3296
  [hep-th]}}.

\bibitem{Susskind:2015toa}
L.~Susskind, ``{The Typical-State Paradox: Diagnosing Horizons with
  Complexity},'' \href{http://dx.doi.org/10.1002/prop.201500091}{{\em Fortsch.
  Phys.} {\bfseries 64} (2016) 84--91},
  \href{http://arxiv.org/abs/1507.02287}{{\ttfamily arXiv:1507.02287
  [hep-th]}}.

\bibitem{deBoer:2018ibj}
J.~de~Boer, R.~Van~Breukelen, S.~F. Lokhande, K.~Papadodimas, and E.~Verlinde,
  ``{On the interior geometry of a typical black hole microstate},''
  \href{http://dx.doi.org/10.1007/JHEP05(2019)010}{{\em JHEP} {\bfseries 05}
  (2019) 010}, \href{http://arxiv.org/abs/1804.10580}{{\ttfamily
  arXiv:1804.10580 [hep-th]}}.

\bibitem{DeBoer:2019yoe}
J.~De~Boer, R.~Van~Breukelen, S.~F. Lokhande, K.~Papadodimas, and E.~Verlinde,
  ``{Probing typical black hole microstates},''
  \href{http://dx.doi.org/10.1007/JHEP01(2020)062}{{\em JHEP} {\bfseries 01}
  (2020) 062}, \href{http://arxiv.org/abs/1901.08527}{{\ttfamily
  arXiv:1901.08527 [hep-th]}}.

\bibitem{Susskind:2020wwe}
L.~Susskind, ``{Black Holes at Exp-time},''
  \href{http://arxiv.org/abs/2006.01280}{{\ttfamily arXiv:2006.01280
  [hep-th]}}.

\bibitem{Harlow:2021dfp}
D.~Harlow and J.-q. Wu, ``{Algebra of diffeomorphism-invariant observables in
  Jackiw-Teitelboim gravity},''
  \href{http://dx.doi.org/10.1007/JHEP05(2022)097}{{\em JHEP} {\bfseries 05}
  (2022) 097}, \href{http://arxiv.org/abs/2108.04841}{{\ttfamily
  arXiv:2108.04841 [hep-th]}}.

\bibitem{AhmedTalkACP}
A.~Almheiri. \url{https://youtu.be/UywgVI62Xvk?t=3934}.
\newblock Talk at Aspen Center for Physics based on work with H.~Verlinde,
  H.~Lin, S.~Collier.

\bibitem{Teitelboim:1983ux}
C.~Teitelboim, ``{Gravitation and Hamiltonian Structure in Two Space-Time
  Dimensions},''
\href{http://dx.doi.org/10.1016/0370-2693(83)90012-6}{{\em Phys. Lett.}
  {\bfseries B126} (1983) 41--45}.

\bibitem{Jackiw:1984je}
R.~Jackiw, ``{Lower Dimensional Gravity},''
\href{http://dx.doi.org/10.1016/0550-3213(85)90448-1}{{\em Nucl. Phys.}
  {\bfseries B252} (1985) 343--356}.

\bibitem{Yang:2018gdb}
Z.~Yang, ``{The Quantum Gravity Dynamics of Near Extremal Black Holes},''
  \href{http://dx.doi.org/10.1007/JHEP05(2019)205}{{\em JHEP} {\bfseries 05}
  (2019) 205},
\href{http://arxiv.org/abs/1809.08647}{{\ttfamily arXiv:1809.08647 [hep-th]}}.

\bibitem{Saad:2019lba}
P.~Saad, S.~H. Shenker, and D.~Stanford, ``{JT gravity as a matrix integral},''
\href{http://arxiv.org/abs/1903.11115}{{\ttfamily arXiv:1903.11115 [hep-th]}}.

\bibitem{Hartman:2013qma}
T.~Hartman and J.~Maldacena, ``{Time Evolution of Entanglement Entropy from
  Black Hole Interiors},''
  \href{http://dx.doi.org/10.1007/JHEP05(2013)014}{{\em JHEP} {\bfseries 05}
  (2013) 014},
\href{http://arxiv.org/abs/1303.1080}{{\ttfamily arXiv:1303.1080 [hep-th]}}.

\bibitem{Susskind:2014rva}
L.~Susskind, ``{Computational Complexity and Black Hole Horizons},''
  \href{http://dx.doi.org/10.1002/prop.201500092}{{\em Fortsch. Phys.}
  {\bfseries 64} (2016) 24--43},
  \href{http://arxiv.org/abs/1403.5695}{{\ttfamily arXiv:1403.5695 [hep-th]}}.
  [Addendum: Fortsch.Phys. 64, 44--48 (2016)].

\bibitem{Kitaev:2018wpr}
A.~Kitaev and S.~J. Suh, ``{Statistical mechanics of a two-dimensional black
  hole},'' \href{http://dx.doi.org/10.1007/JHEP05(2019)198}{{\em JHEP}
  {\bfseries 05} (2019) 198},
\href{http://arxiv.org/abs/1808.07032}{{\ttfamily arXiv:1808.07032 [hep-th]}}.

\bibitem{Harlow:2018tqv}
D.~Harlow and D.~Jafferis, ``{The Factorization Problem in Jackiw-Teitelboim
  Gravity},''
\href{http://arxiv.org/abs/1804.01081}{{\ttfamily arXiv:1804.01081 [hep-th]}}.

\bibitem{Jensen:2016pah}
K.~Jensen, ``{Chaos in AdS$_2$ Holography},''
  \href{http://dx.doi.org/10.1103/PhysRevLett.117.111601}{{\em Phys.\ Rev.\
  Lett.} {\bfseries 117} no.~11, (2016) 111601},
  \href{http://arxiv.org/abs/1605.06098}{{\ttfamily arXiv:1605.06098
  [hep-th]}}.

\bibitem{Maldacena:2016upp}
J.~Maldacena, D.~Stanford, and Z.~Yang, ``{Conformal symmetry and its breaking
  in two dimensional Nearly Anti-de-Sitter space},''
  \href{http://dx.doi.org/10.1093/ptep/ptw124}{{\em PTEP} {\bfseries 2016}
  no.~12, (2016) 12C104}, \href{http://arxiv.org/abs/1606.01857}{{\ttfamily
  arXiv:1606.01857 [hep-th]}}.

\bibitem{engelsoy2016investigation}
J.~Engels{\"o}y, T.~G. Mertens, and H.~Verlinde, ``An investigation of ads 2
  backreaction and holography,'' {\em Journal of High Energy Physics}
  {\bfseries 2016} no.~7, (2016) 139.

\bibitem{Bagrets:2017pwq}
D.~Bagrets, A.~Altland, and A.~Kamenev, ``{Power-law out of time order
  correlation functions in the SYK model},''
  \href{http://dx.doi.org/10.1016/j.nuclphysb.2017.06.012}{{\em Nucl.\ Phys.\
  B} {\bfseries 921} (2017) 727--752},
  \href{http://arxiv.org/abs/1702.08902}{{\ttfamily arXiv:1702.08902
  [cond-mat.str-el]}}.

\bibitem{Cotler:2016fpe}
J.~S. Cotler, G.~Gur-Ari, M.~Hanada, J.~Polchinski, P.~Saad, S.~H. Shenker,
  D.~Stanford, A.~Streicher, and M.~Tezuka, ``{Black Holes and Random
  Matrices},'' \href{http://dx.doi.org/10.1007/JHEP09(2018)002,
  10.1007/JHEP05(2017)118}{{\em JHEP} {\bfseries 05} (2017) 118},
  \href{http://arxiv.org/abs/1611.04650}{{\ttfamily arXiv:1611.04650
  [hep-th]}}.
[Erratum: JHEP09,002(2018)].

\bibitem{Stanford:2017thb}
D.~Stanford and E.~Witten, ``{Fermionic Localization of the Schwarzian
  Theory},'' \href{http://dx.doi.org/10.1007/JHEP10(2017)008}{{\em JHEP}
  {\bfseries 10} (2017) 008}, \href{http://arxiv.org/abs/1703.04612}{{\ttfamily
  arXiv:1703.04612 [hep-th]}}.

\bibitem{naatanen1998weil}
M.~N{\"a}{\"a}t{\"a}nen and T.~Nakanishi, ``Weil-petersson areas of the moduli
  spaces of tori,'' {\em Results in Mathematics} {\bfseries 33} no.~1, (1998)
  120--133.

\bibitem{mirzakhani2007simple}
M.~Mirzakhani, ``Simple geodesics and weil-petersson volumes of moduli spaces
  of bordered riemann surfaces,'' {\em Inventiones mathematicae} {\bfseries
  167} no.~1, (2007) 179--222.

\bibitem{Mathur:2008kg}
S.~D. Mathur, ``{Tunneling into fuzzball states},''
  \href{http://dx.doi.org/10.1007/s10714-009-0837-3}{{\em Gen. Rel. Grav.}
  {\bfseries 42} (2010) 113--118},
  \href{http://arxiv.org/abs/0805.3716}{{\ttfamily arXiv:0805.3716 [hep-th]}}.

\bibitem{Mathur:2008nj}
S.~D. Mathur, ``{Fuzzballs and the information paradox: A Summary and
  conjectures},''
\href{http://arxiv.org/abs/0810.4525}{{\ttfamily arXiv:0810.4525 [hep-th]}}.

\bibitem{Iliesiu:2021ari}
L.~V. Iliesiu, M.~Mezei, and G.~S\'arosi, ``{The volume of the black hole
  interior at late times},''
  \href{http://dx.doi.org/10.1007/JHEP07(2022)073}{{\em JHEP} {\bfseries 07}
  (2022) 073}, \href{http://arxiv.org/abs/2107.06286}{{\ttfamily
  arXiv:2107.06286 [hep-th]}}.

\bibitem{Lin:2018cbk}
H.~W. Lin, ``{Cayley graphs and complexity geometry},''
  \href{http://dx.doi.org/10.1007/JHEP02(2019)063}{{\em JHEP} {\bfseries 02}
  (2019) 063}, \href{http://arxiv.org/abs/1808.06620}{{\ttfamily
  arXiv:1808.06620 [hep-th]}}.

\bibitem{Parker:2018yvk}
D.~E. Parker, X.~Cao, A.~Avdoshkin, T.~Scaffidi, and E.~Altman, ``{A Universal
  Operator Growth Hypothesis},''
  \href{http://dx.doi.org/10.1103/PhysRevX.9.041017}{{\em Phys. Rev. X}
  {\bfseries 9} no.~4, (2019) 041017},
  \href{http://arxiv.org/abs/1812.08657}{{\ttfamily arXiv:1812.08657
  [cond-mat.stat-mech]}}.

\bibitem{Rabinovici:2020ryf}
E.~Rabinovici, A.~S\'anchez-Garrido, R.~Shir, and J.~Sonner, ``{Operator
  complexity: a journey to the edge of Krylov space},''
  \href{http://dx.doi.org/10.1007/JHEP06(2021)062}{{\em JHEP} {\bfseries 06}
  (2021) 062}, \href{http://arxiv.org/abs/2009.01862}{{\ttfamily
  arXiv:2009.01862 [hep-th]}}.

\bibitem{Balasubramanian:2022tpr}
V.~Balasubramanian, P.~Caputa, J.~Magan, and Q.~Wu, ``{Quantum chaos and the
  complexity of spread of states},''
  \href{http://arxiv.org/abs/2202.06957}{{\ttfamily arXiv:2202.06957
  [hep-th]}}.

\bibitem{Saad:2021rcu}
P.~Saad, S.~H. Shenker, D.~Stanford, and S.~Yao, ``{Wormholes without
  averaging},'' \href{http://arxiv.org/abs/2103.16754}{{\ttfamily
  arXiv:2103.16754 [hep-th]}}.

\bibitem{Saad:2021uzi}
P.~Saad, S.~Shenker, and S.~Yao, ``{Comments on wormholes and factorization},''
  \href{http://arxiv.org/abs/2107.13130}{{\ttfamily arXiv:2107.13130
  [hep-th]}}.

\bibitem{Blommaert:2021fob}
A.~Blommaert, L.~V. Iliesiu, and J.~Kruthoff, ``{Gravity factorized},''
  \href{http://arxiv.org/abs/2111.07863}{{\ttfamily arXiv:2111.07863
  [hep-th]}}.

\bibitem{Dray:1984ha}
T.~Dray and G.~'t~Hooft, ``{The Gravitational Shock Wave of a Massless
  Particle},'' \href{http://dx.doi.org/10.1016/0550-3213(85)90525-5}{{\em Nucl.
  Phys. B} {\bfseries 253} (1985) 173--188}.

\bibitem{Shenker:2013pqa}
S.~H. Shenker and D.~Stanford, ``{Black holes and the butterfly effect},''
  \href{http://dx.doi.org/10.1007/JHEP03(2014)067}{{\em JHEP} {\bfseries 03}
  (2014) 067}, \href{http://arxiv.org/abs/1306.0622}{{\ttfamily arXiv:1306.0622
  [hep-th]}}.

\bibitem{thurston1979geometry}
W.~P. Thurston and J.~W. Milnor, ``The geometry and topology of
  three-manifolds,'' 1979.
\newblock \url{http://library.msri.org/books/gt3m/}.

\bibitem{Witten:2020ert}
E.~Witten, ``{Deformations of JT Gravity and Phase Transitions},''
  \href{http://arxiv.org/abs/2006.03494}{{\ttfamily arXiv:2006.03494
  [hep-th]}}.

\bibitem{Maxfield:2020ale}
H.~Maxfield and G.~J. Turiaci, ``{The path integral of 3D gravity near
  extremality; or, JT gravity with defects as a matrix integral},''
  \href{http://dx.doi.org/10.1007/JHEP01(2021)118}{{\em JHEP} {\bfseries 01}
  (2021) 118}, \href{http://arxiv.org/abs/2006.11317}{{\ttfamily
  arXiv:2006.11317 [hep-th]}}.

\end{thebibliography}\endgroup

\bibliographystyle{utphys}

\end{document}